\providecommand{\ReferenceMode}{0}
\ifnum\ReferenceMode=0
  \newcommand{\ReferenceStyle}{aasjournal}
\else
  \newcommand{\ReferenceStyle}{aasjournal}
\fi

\let\originalbibliographystyle\bibliographystyle
\renewcommand{\bibliographystyle}[1]{\originalbibliographystyle{\ReferenceStyle}}
\documentclass[usenatbib,useAMS,onecolumn]{aastex63}
\let\bibliographystyle\originalbibliographystyle

\makeatletter
\ifnum\ReferenceMode=0
  \setcitestyle{authoryear,round,semicolon,aysep={},yysep={,}}
  \def\NAT@sort{0}
  \def\NAT@cmprs{0}
\else
  \setcitestyle{numbers,square,comma}
  \def\NAT@sort{1}
  \def\NAT@cmprs{1}
\fi
\makeatother
\makeatletter
\renewcommand{\paragraph}{%
  \@startsection{paragraph}{4}{\z@}%
  {2ex plus 1ex minus .2ex}%
  {1ex plus .2ex}%
  {\normalfont\small\itshape\center}}
\makeatother

\usepackage[T1]{fontenc}
\usepackage{graphicx}
\usepackage{amsmath,amssymb,amscd}
\usepackage{hyperref, multirow, aas_macros, makecell, verbatim} %%dcolumn,

\usepackage{acronym}
\usepackage{CJK}
\usepackage{ulem, cancel} % for the purpose of highlight changes, will remove after modifying

\newcommand{\erg}{\,\mathrm{erg}}

\newcommand{\g}{\,\mathrm{g}}
\newcommand{\s}{\,\mathrm{s}}
\newcommand{\cm}{\,\mathrm{cm}}
\newcommand{\km}{\,\mathrm{km}}
\def\apjl{Astrophys. J. Lett.}
\def\apjs{Astrophys. J. Suppl.}

\def\aap{Astron. Astrophys.}

\usepackage{lineno}
\let\oldequation\equation
\let\oldendequation\endequation
\renewenvironment{equation}{\linenomathNonumbers\oldequation}{\oldendequation\endlinenomath}

\let\oldalign\align
\let\oldendalign\endalign
\renewenvironment{align}{\linenomathNonumbers\oldalign}{\oldendalign\endlinenomath}

\let\oldgather\gather
\let\oldendgather\endgather
\renewenvironment{gather}{\linenomathNonumbers\oldgather}{\oldendgather\endlinenomath}

\newcommand{\DOA}{Department of Astronomy, University of Science and Technology of China, Hefei, Anhui 230026, China.
}

\newcommand{\SASS}{School of Astronomy and Space Sciences, University of Science and Technology of China, Hefei, Anhui 230026, China.}
\newcommand{\GZU}{College of Physics, Guizhou University, Guiyang, Guizhou 550025, China.}
\newcommand{\IHEP}{Key Laboratory for Particle Astrophysics, Institute of High Energy Physics, Chinese Academy of Sciences, Beijing 100049, China.}
\newcommand{\CoAut}{The co-first authors (He, Zhu, Chen) contribute equally to this review.}
\newcommand{\CorAut}{Corresponding author: wzhao7@ustc.edu.cn}

\renewcommand\thetable{\Roman{table}}
\newcommand{\D}{\mathrm{d}}

\begin{document}

%\linenumbers

\begin{CJK*}{UTF8}{gbsn} % Use default fonts from CJK (see below)

%%%================================================
\title{Electromagnetic Counterparts of Stellar-mass Binary Black Hole Mergers in AGN Accretion Disks: Theoretical Models and Observational Status}

%%%=============================================

\author{Lei He 
%\orcidlink{0000-0001-7613-5815}
}
\thanks{\CoAut}
\affiliation{\DOA}
\affiliation{\SASS}

\author{Liang-Gui Zhu
%\orcidlink{0000-0001-7688-6504}
}
\thanks{\CoAut}
\affiliation{\DOA}
\affiliation{\SASS}

\author{Ken Chen 
%\orcidlink{XXX}
}
\thanks{\CoAut}
\affiliation{\DOA}
\affiliation{\SASS}

\author{Zi-Gao Dai
%\orcidlink{XXX}
}
%\thanks{\CoAut}
\affiliation{\DOA}
\affiliation{\SASS}

\author{Ye-Fei Yuan
%\orcidlink{XXX}
}
%\thanks{\CoAut}
\affiliation{\DOA}
\affiliation{\SASS}

\author{Jian-Min Wang 
%\orcidlink{XXX}
}
%\thanks{\CoAut}
\affiliation{\IHEP}

\author{Wen Zhao 
%\orcidlink{0000-0002-1330-2329}
}
\thanks{\CorAut}
\affiliation{\DOA}
\affiliation{\SASS}
\affiliation{\GZU}

%%%====================================================
%%%====================================================
\begin{abstract}

The discovery of the binary neutron star merger GW170817 and its electromagnetic counterparts marked the beginning of the era of multimessenger gravitational-wave astronomy. In contrast, binary black hole (BBH) mergers, which constitute the majority of gravitational-wave sources detected to date, are generally expected to be electromagnetically dark. However, if BBH mergers occur within the dense gaseous environments of active galactic nucleus (AGN) accretion disks, interactions between the merging black holes and surrounding gas may produce electromagnetic emission across multiple wavelengths. AGN disks may therefore provide a natural multimessenger laboratory for studying BBH mergers and their astrophysical environments. A key question is whether such a scenario is supported by current observations and whether the associated electromagnetic signatures are sufficiently distinctive and detectable. In this review, we summarize the current theoretical and observational status of BBH mergers in AGN accretion disks, focusing on the evidence for their occurrence, the physical mechanisms and expected signatures of electromagnetic emission, and searches for candidate counterparts. We also discuss the major challenges in identifying and confirming these associations and the prospects for future multimessenger observations with current and forthcoming multiwavelength facilities.

\end{abstract}

\keywords{Gravitational-wave detection, Binary black hole population, Active galactic nucleu, Electromagnetic counterpart }

%%%====================================================

\section{Introduction}
\label{intro}

In 2015, the Laser Interferometer Gravitational-Wave Observatory (LIGO) made the first direct detection of a binary black hole (BBH) merger, GW150914, ushering in the era of gravitational-wave (GW) astronomy \citep{abbottObservationGravitationalWaves2016}. Since then, the international ground-based GW detector network, LIGO-Virgo-KAGRA (LVK), has continuously detected hundreds of compact-binary coalescences through multiple observing runs, including binary neutron star (BNS), neutron star-black hole (NSBH), and BBH mergers. GW signals provide direct measurements of key properties of compact binaries, including luminosity distance, sky localization, component masses and spins, and orbital dynamics, offering unique opportunities to study compact-object astrophysics, test gravity in the strong-field regime, and probe cosmology. However, GW waveforms alone provide limited information about the physical environments surrounding the sources, such as gas density, magnetic fields, and radiation fields, as well as their coupling to the merger process. This limitation leads to substantial uncertainties in our understanding of the formation channels, radiation mechanisms, and host galaxies of GW sources. Multi-messenger observations therefore provide a crucial avenue for overcoming these limitations. Joint observations of GWs and electromagnetic (EM) radiation, together with neutrinos and cosmic rays when available, can provide complementary information on the dynamics, environment, and radiation of compact-object mergers. The landmark detection of GW170817 and its broadband EM counterparts \citep{abbottMultimessengerObservationsBinary2017,abbottGW170817ObservationGravitational2017,abbottGravitationalWavesGammaRays2017} demonstrated the transformative power of this approach: EM observations can precisely identify the host galaxy and determine its redshift, while multiwavelength observations can constrain the jet structure, ejecta mass and composition, thereby providing critical insights into the origin of heavy elements, the neutron-star equation of state, and cosmological parameters. GW astronomy has thus entered the era of multi-messenger astrophysics, which has become one of the most frontier areas of modern astronomy and physics. Nevertheless, multi-messenger GW astronomy remains in a stage of extremely limited confirmed samples. By the end of LVK O4, only one GW event had a firmly established EM counterpart, and this severe scarcity of confirmed associations has become a major bottleneck for further progress in the field.

At present, systematic searches for EM counterparts to GW events have focused primarily on BNS and NSBH mergers \citep{a1,a2,a3,a4,a5,a6,a7}. Their associated EM emission is relatively well understood: tidal disruption, dynamical ejecta, and relativistic jets can give rise to gamma-ray bursts, kilonovae, and broadband afterglows, respectively, providing relatively clear temporal and spectral signatures to guide follow-up observations. However, these events have relatively low occurrence rates, and their sky localizations can span tens to thousands, or even more than ten thousand, square degrees when the detector network and sensitivity are limited. This makes rapid, deep, and multiwavelength coverage extremely challenging. Moreover, EM detectability is strongly affected by viewing angle, dust extinction, and background contamination, making observable counterparts intrinsically rare and consequently limiting our ability to investigate their radiation mechanisms and formation channels.

In contrast, BBH mergers account for more than 90\% of the currently detected GW events and constitute the most abundant and statistically promising source population \citep{abacOpenDataLIGO2026}. Yet their EM counterparts have remained elusive. The conventional picture assumes that BBH mergers in vacuum contain little or no radiating matter and therefore produce negligible EM emission, making BBH mergers a less obvious target for multi-messenger studies for many years. Recent theoretical developments, however, suggest that this picture may not be universally valid. If a BBH merger occurs in a dense gaseous environment, particularly within an active galactic nucleus (AGN) accretion disk, interactions between the binary black holes and the surrounding gas before and after merger, the recoil of the remnant black hole and its perturbation of the disk, as well as shocks, magnetic reconnection, and jet formation, may generate detectable broadband EM emission from radio and optical wavelengths to the high-energy regime.

AGN accretion disks are generally considered highly active environments for star formation and stellar evolution \citep{1999ApJ...521..502C}. This picture is supported by several lines of indirect observational evidence. For example, spectroscopic studies have found little evolution of quasar metallicity with cosmological redshift, suggesting intense and recurrent stellar activity within AGN disks \citep{nagaoEvolutionBroadlineRegion2006,wangMetallicityQuasarBroadline2022}. Observations have also suggested that the gamma-ray burst GRB191019A may have originated from a binary neutron star merger in an AGN environment \citep{levanLongdurationGammarayBurst2023,lazzatiGRB191019AShort2023}. More recently, the discovery of quasi-periodic eruptions \citep{miniuttiNinehourXrayQuasiperiodic2019}, if interpreted as arising from interactions between orbiting compact objects and the surrounding disk gas, provides further evidence that AGN accretion disks may host abundant populations of stars and compact objects. The deaths of massive stars produce large numbers of compact remnants, some of which may subsequently form compact binaries and potentially contribute significantly to the BBH population observed by LVK. Nevertheless, owing to the complexity of the accretion-disk environment, theoretical models for EM emission from BBH mergers in AGN disks remain diverse and actively debated. Depending on the underlying central engine, these models can be broadly divided into two categories: recoil-driven models, including shock breakout from dynamically stripped gas \citep{mckernanRampressureStrippingKicked2019}, shock-heated wake emission \citep{grahamCandidateElectromagneticCounterpart2020}, outflow shock breakout \citep{kimuraOutflowBubblesCompact2021}, cooling emission from outflow shocks \citep{rodriguez-ramirezOpticalUVFlares2025}, thermal emission from jet-cocoon systems \citep{chenElectromagneticCounterpartsPowered2024}, and cocoon cooling emission \citep{rodriguez-ramirezOpticalEmissionModel2023}; and in-situ accretion models, including jet cooling emission and Bondi-like accretion-powered flares \citep{wangAccretionmodifiedStarsAccretion2021}, jet-shock breakout followed by non-thermal emission \citep{tagawaHighenergyElectromagneticNeutrino2023}, and cocoon-shock cooling emission \citep{tagawaShockCoolingBreakout2024}.

Meanwhile, several environment-sensitive signatures have been identified in the observed GW population, including unusually massive black holes, extreme mass ratios, high spins, and the possible nonzero orbital eccentricities. Statistical studies have reported significant spatial correlations between GW events and AGN populations, as well as between GW events and AGNs exhibiting anomalous flares \citep{zhuEvidenceFractionLIGO2025,zhuConstrainingFractionLIGO2026}. These findings provide statistical support for the possibility that a fraction of BBH mergers may originate from dynamical or accretion-disk-assisted formation channels. In particular, observations by Zwicky Transient Facility (ZTF), Dark Energy Camera (DECam), Wide Field Survey Telescope (WFST), and other facilities have identified candidate AGN flares potentially associated with several BBH events, including the most massive BBH mergers detected to date, GW190521, as well as GW231123, GW190803, and S230922g \citep{grahamCandidateElectromagneticCounterpart2020,grahamLightDarkSearching2023,cabreraSearchingElectromagneticEmission2024,heTracingLightIdentification2025,heSearchingElectromagneticCounterpart2026,zhangLVKS241125nMassive2026}. These intriguing associations have stimulated extensive discussions of BBH merger mechanisms in AGN disks and the statistical significance of GW-AGN correlations \citep{mckernanRampressureStrippingKicked2019,yangHierarchicalBlackHole2019,heSystematicSearchActive2025}, bringing BBH-AGN multi-messenger astrophysics into increasing international focus.

Despite these rapid theoretical and observational developments, research in this field has so far focused predominantly on model building and numerical simulations, whereas systematic analyses of real observational data remain comparatively underdeveloped. In particular, searches for BBH EM counterparts have largely remained at the level of individual candidate events, without a systematic framework for identification and confirmation. Fundamental questions therefore remain unanswered: How can BBH EM counterparts be robustly identified from real multiwavelength observations? How can EM counterparts be used to constrain or distinguish among competing emission models? And can a statistically significant sample of identified counterparts be used to probe the physics of AGN accretion disks themselves?

These challenges arise primarily because, although the AGN-disk scenario provides a physically plausible environment for producing EM counterparts to BBH mergers, robust observational identification remains difficult. First, AGNs exhibit stochastic variability on timescales ranging from days to years, arising from processes such as stochastic accretion variability, disk instabilities, tidal disruption events (TDEs), and nuclear supernovae. Without additional constraints, a single flare cannot readily be attributed to a GW merger, resulting in a potentially very high false-association rate. Second, the sky localization regions of BBH events typically span tens to thousands of square degrees and can contain a very large number of AGNs, making spatial overlap alone insufficient to establish a credible physical association. Third, different theoretical models predict different delays, durations, spectral properties, and recurrence behavior of the associated EM signals \citep{mckernanRampressureStrippingKicked2019,grahamCandidateElectromagneticCounterpart2020,kimuraOutflowBubblesCompact2021,rodriguez-ramirezOpticalUVFlares2025,chenElectromagneticCounterpartsPowered2024,rodriguez-ramirezOpticalEmissionModel2023,wangAccretionmodifiedStarsAccretion2021,tagawaHighenergyElectromagneticNeutrino2023,tagawaShockCoolingBreakout2024}. These differences provide valuable diagnostic information, but also make the systematic search and robust identification of counterparts particularly challenging. Addressing these issues requires a unified observational framework that integrates gravitational-wave information with multiwavelength time-domain observations and physically motivated models of AGN-associated electromagnetic signatures, paving the way toward a systematic understanding of the origin, nature, and diversity of EM counterparts to BBH mergers and emerging as a major research frontier in the coming years.

In this review, we provide a systematic overview of the major advances in this rapidly developing field, with particular emphasis on three key questions. First, do the gravitational-wave observations accumulated by the LVK to date provide evidence that at least a fraction of BBH mergers originate in AGN accretion disks? We discuss this question from two complementary perspectives: whether the observed source-parameter distributions exhibit signatures consistent with an AGN-disk origin, and whether the spatial association between BBH mergers and AGNs lends independent support to this scenario. These observational constraints are reviewed in Section 2. Second, from a theoretical perspective, if BBH mergers do indeed occur within AGN accretion disks, can they generate sufficiently luminous electromagnetic emission to be detectable across multiple wavelengths? We examine the expected spectral and temporal properties of such emission, including its distribution across different wavebands, characteristic duration and temporal evolution, and the possible time delay between the GW merger and the associated EM signal. These theoretical predictions are discussed in Section 3. Finally, Section 4 reviews current efforts to identify EM counterparts to BBH mergers through multiwavelength observations, with particular emphasis on optical searches. We summarize the current observational status, discuss promising strategies for confirming candidate counterparts through coordinated multiwavelength observations, and highlight the broader astrophysical and cosmological opportunities that may arise once such counterparts are securely identified. Section 5 is devoted to a summary and outlook of this article.

\section{Observational Constraints on the AGN Formation Channel}
\subsection{Bayesian inferences of the BBH population properties}
The detection of GW150914 and subsequent BBH merger GW events transformed black-hole astrophysics \citep{abbottPropertiesBinaryBlack2016}. 
The observation established not only the existence of binary black-hole mergers 
but also the existence of a population of stellar-mass black holes 
substantially heavier than many black holes then known from X-ray binaries \citep{abbottASTROPHYSICALIMPLICATIONSBINARY2016}. 
The subsequent observing runs rapidly changed the problem: instead of asking whether BBHs exist, 
the community could ask how their masses, mass ratios, spins, merger rates, 
and cosmic evolution are distributed \citep{abbottBinaryBlackHole2019,abbottPopulationPropertiesCompact2021,abbottPopulationMergingCompact2023,abacGWTC40PopulationProperties2026,abacGWTC50PopulationProperties2026}.
The cumulative GWTC-1 through GWTC-5.0 \citep{abbottGWTC1GravitationalWaveTransient2019a,abbottGWTC2CompactBinary2021,abbottGWTC3CompactBinary2023,abbottGWTC21DeepExtended2024,abacGWTC40UpdatingGravitationalwave2026,abacGWTC50ObservationsSecond2026}
sequence illustrates this transition particularly well. 

The primary-mass spectrum: 
The dominant feature of the current population is that the primary-mass distribution is not a featureless power law. 
Early GWTC-1 analyses were adequately summarized by a truncated power law \citep{abbottBinaryBlackHole2019}, 
but the growing sample revealed local structure. GWTC-4.0 and GWTC-5.0 both identify 
a robust enhancement around 10 $M_\odot$ and a change in slope around 35 $M_\odot$, 
with weaker evidence for a feature near 20 $M_\odot$ \citep{abacGWTC40PopulationProperties2026,abacGWTC50PopulationProperties2026}. 

The mass-ratio distribution: 
The mass-ratio distribution has evolved from being effectively unconstrained to a measurable component of the population model. 
GWTC-4.0 reported a peak near $q \equiv m_2/m_1 = 0.74$ for systems with primary masses around 10 $M_\odot$, which can be qualitatively consistent with stable mass transfer in isolated binaries \citep{abacGWTC40PopulationProperties2026}. 
GWTC-5.0 finds reduced evidence for that particular $q\simeq 0.7$ feature and 
instead places more weight on approximately equal masses overall, while simultaneously finding that systems 
with primary masses above about 40 $M_\odot$ favor unequal-mass binaries \citep{abacGWTC50PopulationProperties2026}. 
The changing inference is a useful reminder that population conclusions depend on both 
sample size and the flexibility of the population model. 

The spin distribution: 
The dimensionless spin magnitude $\chi$ is now measured well enough for population statements, 
but individual spins remain challenging because the inspiral signal is much more sensitive to 
particular combinations of spin components than to all three-dimensional 
spin-vector degrees of freedom \citep{santamariaMatchingPostNewtonianNumerical2010,ajithInspiralMergerRingdownWaveformsBlackHole2011}. 
GWTC-4.0 and GWTC-5.0 both favor predominantly non-extremal spins \citep{abacGWTC40UpdatingGravitationalwave2026,abacGWTC50ObservationsSecond2026}, 
and GWTC-5.0 estimates that roughly $69\%-84\%$ of black holes have $\chi \lesssim 0.5$ in its fiducial analyses. 
Yet the effective inspiral spin, $\chi_{\rm eff}$, which describes 
the mass-weighted projection of the component spins parallel to the orbital angular momentum and is defined as 
\[ \chi_{\rm eff} \equiv \frac{m_1 \chi_1 \cos\theta_1 + m_2 \chi_2 \cos\theta_2}{m_1 + m_2}, \]
where $\theta_{1,2}$ are the angles between spins and orbital angular momentum, is strongly informative about formation. 
The distribution peaks near zero but is asymmetric: a substantial fraction of binaries 
have negative $\chi_{\rm eff}$, while a smaller but nonzero fraction shows evidence for 
a preference toward positive, aligned spins \citep{abacGWTC50PopulationProperties2026}. 

The merger rate and redshift evolution: 
GWTC-5.0 infers a BBH merger-rate density of $27.5 - 49.4 \, \mathrm{Gpc}^{-3}\,\mathrm{yr}^{-1}$ at $z=0.2$ 
for component masses in the $2.5 - 200 \, M_\odot$ mass range. 
More importantly for formation-channel work, the selection-corrected population can be tested 
for redshift-dependent changes in mass and spin distributions. 
GWTC-5.0 reports evidence that the width of the $\chi_{\rm eff}$ distribution broadens with redshift, 
although details depend on the adopted model. Such evolution could reflect changing metallicity, 
changing mixture fractions of formation channels, 
or different delays between binary formation and merger \citep{abacGWTC50PopulationProperties2026}. 

% %%%----------------------------------------------------
\subsubsection{Candidate BBH formation channels}

The astrophysical environments capable of generating the BBH population detected by LVK remain an open question to date.
The LVK population papers generally organize BBH formation into 
two broad classes: isolated binary evolution \citep{postnovEvolutionCompactBinary2014}
and dynamical formation \citep{mapelliHierarchicalBlackHole2021,mapelliCosmicEvolutionBinary2022}.
The dynamical class contains several physically different environments, 
a practical classification is \citep{mandelMergingStellarmassBinary2022,mandelRatesCompactObject2022}:  
(i) dynamical assembly in globular, young, nuclear, or other star clusters; 
(ii) hierarchical merger, usually as a secondary-generation process embedded in a dynamical environment; and 
(iii) AGN formation channel \citep{yangHierarchicalBlackHole2019,yangAGNDisksHarden2019,tagawaFormationEvolutionCompactobject2020}
including dynamical assemblies and hierarchical mergers. 

Isolated binary evolution \citep{postnovEvolutionCompactBinary2014}: 
two massive stars are born as a binary and evolve together through wind mass loss, 
tides, Roche-lobe overflow, stable or unstable mass transfer, common-envelope evolution, and compact-object formation. 
The two black holes therefore inherit correlated masses and, in many models, correlated spin directions. 
The natural advantages of this channel are high rates in ordinary star formation and the ability to produce strong spin-orbit alignment when tidal coupling is efficient. 
A major weakness is that highly massive systems in the pair-instability range and repeated-merger products 
are difficult to generate directly, while strong kicks and common-envelope uncertainties can 
suppress the merger rate \citep{belczynskiEffectPairinstabilityMass2016,woosleyPairinstabilityMassGap2021,giacobboProgenitorsCompactobjectBinaries2018}. 

Dynamical assembly \citep{rodriguezBinaryBlackHole2015,rodriguezBinaryBlackHole2016b} and hierarchical merger \citep{kimballEvidenceHierarchicalBlack2021,liuHierarchicalBlackHole2021}: 
in a dense stellar system, black holes can interact through binary-single and binary-binary encounters, exchange companions, harden binaries, and eventually merge. 
The orbital orientation of a newly formed binary is largely decoupled from the progenitor stellar spins, 
producing approximately isotropic spin tilts. 
This makes negative $\chi_{\rm eff}$ and precession natural outcomes \citep{tagawaSignaturesHierarchicalMergers2021}. 
Cluster dynamics also provide a mechanism for retaining merger remnants and assembling hierarchical binaries, 
provided the escape speed is sufficiently high or GW recoil is not too large \citep{islamKickMattersImpact2026}. 
Hierarchical mergers are efficient in environments with a high density of black holes and sufficient retention, 
including young clusters, massive clusters, nuclear star clusters \citep{mapelliHierarchicalBlackHole2021,mapelliCosmicEvolutionBinary2022}, 
and AGN disks. 

AGN formation channel \citep{yangHierarchicalBlackHole2019,yangAGNDisksHarden2019,tagawaFormationEvolutionCompactobject2020}: 
an active galactic nucleus contains a massive accreting black hole surrounded by a dense gaseous disk. 
Stellar-mass black holes embedded in the disk can lose orbital energy to gas, become geometrically confined 
toward the disk midplane, migrate radially, accumulate in migration traps, and encounter other black holes. 
Gas torques can promote binary formation and hardening, while accretion can alter masses, spins, 
and merger rates \citep{yangAGNDisksHarden2019,tagawaFormationEvolutionCompactobject2020,tagawaSpinEvolutionStellarmass2020,bertiInferringBlackHole2026}.

% %%%----------------------------------------------------
\subsubsection{Signature of AGN formation channel}

An AGN disk is qualitatively different from a purely stellar dynamical system. 
Gravity, gas drag, radiation, and accretion all act simultaneously \citep{yangHierarchicalBlackHole2019,yangAGNDisksHarden2019,
tagawaFormationEvolutionCompactobject2020, tagawaSpinEvolutionStellarmass2020,mckernanMcFACTSTestingLVK2025,vaccaroAGNdrivenBBHMergers2026}.
Stellar-mass black holes can be supplied to the nuclear region by the surrounding stellar population, 
embedded black holes can interact with the disk, and a dense gas reservoir can dissipate orbital energy. 
These effects create a natural pipeline: capture or formation of black holes in the disk $\to$ 
migration $\to$ repeated encounters $\to$ binary formation $\to$ gas-assisted hardening $\to$ 
merger $\to$ remnant retention $\to$ possible re-entry into the merger cycle. 

The strongest conceptual advantage of the AGN formation channel is its ability to 
efficiently assemble binaries and recycle merger remnants. 
If the remnant is retained in the disk, it can merge again quickly. 
A sequence of such mergers can populate masses that are difficult or impossible to obtain from single-star collapse, 
including the region between the upper stellar-remnant mass scale and the conventional pair-instability mass gap. 
Hierarchical assembly can also generate high spin magnitudes. 
A nearly non-spinning pair of first-generation black holes can produce a merger remnant with 
a substantial spin because the binary orbital angular momentum is converted into the remnant's angular momentum. 
Additionally, gas accretion onto BHs can also increase the spin angular momentum, 
leading to larger spins for first-generation BBHs and driving the spins of higher-generation BBHs 
toward unity \citep{tagawaSpinEvolutionStellarmass2020,bertiInferringBlackHole2026, vaccaroAGNdrivenBBHMergers2026}. 

Another important feature of the dynamics of BBH formation in AGN disks and in other environments is that 
gas torques can align the angular momentum of an embedded black hole or binary with 
the disk angular momentum \citep{yangHierarchicalBlackHole2019, yangAGNDisksHarden2019, 
tagawaFormationEvolutionCompactobject2020, tagawaSpinEvolutionStellarmass2020, mckernanMcFACTSTestingLVK2025, bertiInferringBlackHole2026, vaccaroAGNdrivenBBHMergers2026}. 
This creates a distinctive possibility: a dynamically assembled system can still show preferential spin alignment. 
That combination is important because purely gas-free dynamical assembly naturally predicts nearly isotropic orientations, 
while isolated binaries more naturally produce alignment because the stars share an evolutionary history. 
AGN disks can therefore occupy an intermediate but diagnostically useful region of parameter space: dynamically assembled binaries with a tendency toward alignment. 

AGN disks can, in principle, overcome one of the main limitations of ordinary dynamical environments: the need for a sufficiently high black-hole interaction rate in a small volume. A galactic nucleus can host a comparatively large reservoir of black holes, while the disk can concentrate them into a geometrically thin, dense region. This can increase repeated encounters and promote merger recycling. In addition, AGN activity occurs over cosmic time and in galaxies where gas supply and metallicity evolve, potentially linking BBH merger-rate evolution to the cosmic history of black-hole accretion \citep{vaccaroHierarchicalBlackHole2025}.

The above discussion concerns the qualitative expectations for the BBH population produced through the AGN formation channel. However, quantitative modeling of the BBH population is considerably more challenging. 
Owing to the intrinsic complexity of AGN disks, the interactions between black holes and the surrounding gas, 
and the dynamical interactions among black holes, it is difficult to construct fully reliable analytical models. 
Consequently, quantitative predictions for the population properties of AGN-channel BBHs have largely relied on numerical simulations.

Early simulation studies adopted relatively simplified astrophysical assumptions and focused primarily on the properties of BBHs formed through successive generations of mergers, with representative work presented in \citet{yangHierarchicalBlackHole2019}. 
Subsequent studies incorporated more realistic AGN environments, and began to investigate both 
the BBH merger efficiency and the dependence of the resulting BBH population on different AGN models 
\citep{yangAGNDisksHarden2019, tagawaFormationEvolutionCompactobject2020, tagawaSpinEvolutionStellarmass2020}. 
Studies such as \citet{bellovaryMIGRATIONTRAPSDISKS2016, pengLastMigrationTrap2021} and \citet{gonglewskiOrbitalMigrationInteracting2026} further 
incorporated the migration of stellar-mass black holes within the AGN disk, 
thereby providing a more self-consistent treatment of their spatial evolution and merger dynamics. 
More recently, numerical frameworks for modeling BBH formation and evolution in AGN disks have developed into 
publicly available software packages, with representative examples including \citet{mckernanMcFACTSTestingLVK2025, cookMcFACTSIIMass2025, delfaveroMcFACTSIIICompact2025} and \citet{vaccaroAGNdrivenBBHMergers2026}. 
As an increasing number of relevant astrophysical processes and dynamical effects are incorporated, 
these simulations have progressively become more sophisticated and increasingly realistic representations of BBH formation in AGN environments.

Synthesizing the relevant studies, it is evident that, owing to the complexity of BBH formation in the AGN channel, 
the AGN channel does not predict a unique set of one-dimensional distributions for binary black hole parameters. 
Rather, it generically predicts a population shaped by the interplay between gas capture, migration, 
accretion, dynamical encounters, spin alignment, and hierarchical mergers. 
Nevertheless, at a qualitative level, BBH populations formed through the AGN channel 
are expected to exhibit the following characteristic features: 
\begin{itemize}
  \item High-mass tail in the mass distribution: produced by preferential capture and, especially, repeated mergers of stellar-mass black holes. 
  \item Low-mass-ratio subpopulation: repeated mergers between remnants of different generations also 
        tend to produce unequal-mass binaries. 
  \item High-spin subpopulation: hierarchical assembly naturally generates massive remnants with 
        spin magnitudes of order $\chi \sim 0.7$, while efficient gas accretion can further spin black holes 
        up toward $\chi \sim 1$. 
  \item Spin-alignment-dependent $\chi_{\rm eff}$ distributions: the AGN channel can generate 
        a broad distribution of $\chi_{\rm eff}$ centered near zero when binary--single interactions randomize 
        the binary orbital angular momentum, despite relatively high individual spin magnitudes. 
        Conversely, efficient gas accretion and spin alignment can produce preferentially 
        positive $\chi_{\rm eff}$. 
  \item Candidate correlations among mass, mass ratio, and spin parameters: 
        hierarchical mergers favor high mass and low mass-ratio systems, leading to a negative $m_1-q$ correlation; 
        while efficient accretion and spin alignment can additionally produce high $|\chi_{\rm eff}|$ systems, 
        leading to a negative $q-\chi_{\rm eff}$ correlation and a positive $m_1-\chi_{\rm eff}$ correlation. 
\end{itemize}

Consequently, the most discriminating signature of the AGN channel is unlikely to be any single parameter, 
but rather a correlated high-dimensional structure involving mass, mass ratio, spin magnitude, spin alignment, 
and, potentially, orbital eccentricity \citep{zevinImplicationsEccentricObservations2021, samsingAGNPotentialFactories2022, roznerUniversalEccentricityDistribution2026}.

% %%%----------------------------------------------------
\subsubsection{Comparison of individual BBH properties with AGN channel predictions}

The difficulties encountered by isolated binary evolution and dynamical-assembly and hierarchical-merger scenarios in non-AGN environments when attempting to reproduce BBH mergers with extreme properties, 
such as GW190521 \citep{abbottGW190521BinaryBlack2020} and GW231123 \citep{abacGW231123BinaryBlack2025}, 
have increased the interest in the AGN formation channel. 
Importantly, the issue is not that alternative channels are fundamentally incapable of producing such systems, but rather that reproducing several extreme properties simultaneously may require relatively restricted combinations of progenitor and environmental parameters. 
AGN disks provide an alternative dynamical environment in which repeated mergers, gas-assisted binary evolution, and the retention of merger remnants can operate simultaneously, potentially allowing massive and hierarchical BBHs to form without requiring all of the extreme properties to be present in the first-generation BH population.

For GWTC-1 BBH events and the specific GW190521 event, 
\citet{liComparingHierarchicalBlack2023} compared hierarchical mergers in AGN disks with those 
in stellar clusters and examined the resulting distributions of mass, mass ratio, effective spin, 
and precession-related spin parameters. 
They found that hierarchical mergers in AGN disks preferentially populate higher primary masses than those in stellar clusters, 
with the characteristic peak of the primary-mass distribution around $50\,M_\odot$, compared with approximately $13\,M_\odot$ for their cluster model. 
The AGN population also exhibited a relatively narrow and positively shifted $\chi_{\rm eff}$ distribution, with a peak at $\chi_{\rm eff}\gtrsim0.3$, whereas the cluster population was approximately symmetric around zero. 
Their analysis further showed that the mass-ratio and effective-precession-spin distributions can help distinguish AGN and stellar-cluster hierarchical mergers. 
When these predictions were compared with the LVK hierarchical-merger candidates, 
the authors argued that a substantial fraction, potentially even the majority, 
of the hierarchical-merger candidates could be of AGN origin if AGN disks account for a sufficiently large fraction of the overall hierarchical-merger rate. 
This result therefore showed that the subset of BBHs with evidence for hierarchical assembly can be consistent with an AGN contribution, although the same signatures can arise in other dense dynamical environments and the conclusion remains dependent on the adopted population models and merger-rate assumptions.

For GW231123, the most massive BBH mergers reported by the LVK to date, 
in principle, both isolated binary evolution and hierarchical mergers in dense stellar environments 
may produce systems with comparable masses, particularly if the progenitor population 
is assumed to have very low metallicity \citep{tanikawaGW231123FormationPopulation2026,liuFormationGW231123Population2025,angeloniInvestigatingFormationChannel2026}. 
However, such scenarios generally rely on specific assumptions about the initial BH mass spectrum and the efficiency of hierarchical growth \citep{liHierarchicalMergerScenario2025, liGW231123LikelyProduct2026,passengerGW231123HierarchicalMerger2026}. 
By contrast, \citet{delfaveroProspectsFormationGW2311232025} investigated BBH formation in AGN disks using the \textsf{McFACTS} Monte Carlo simulation code and found that GW231123-like systems can be produced through hierarchical mergers over a range of model assumptions, with the extreme component masses arising naturally from repeated merger generations rather than requiring exceptionally massive first-generation BHs. 
In this sense, the AGN channel provides a physically plausible pathway to the observed mass scale because the deep nuclear potential and gas-assisted binary evolution can facilitate the retention, migration, and subsequent re-merger of previous merger remnants.

\begin{figure}[htbp] 
 \centering
 \includegraphics[width=0.7\textwidth]{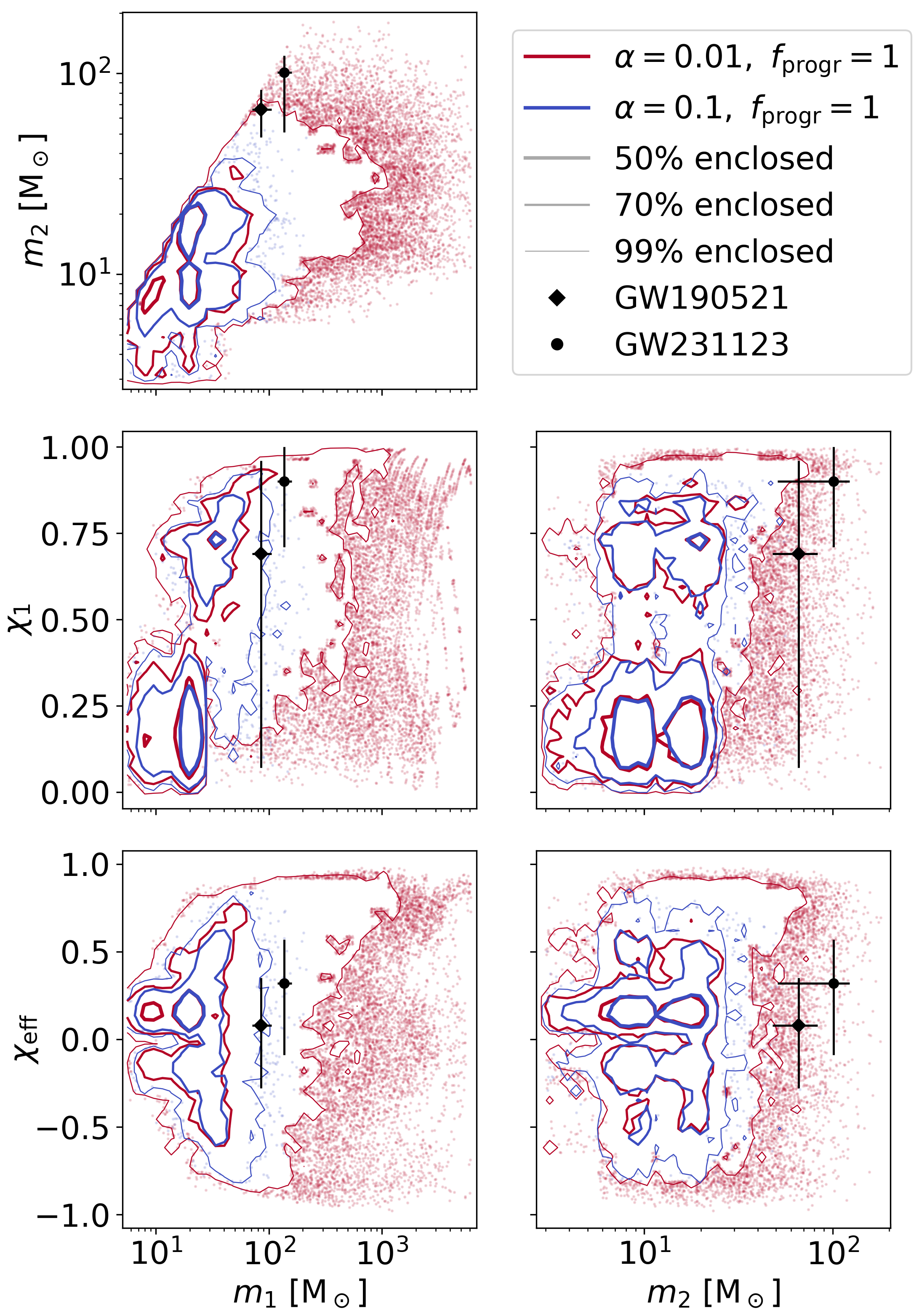}
 \caption{Parameter distributions of mock BBH events for a simulated AGN population \citep{vaccaroAGNdrivenBBHMergers2026}, 
 comparing two disk viscosities $\alpha = 0.01$ (red) and $\alpha = 0.1$ (blue), 
 both for fully prograde configurations $f_{\rm progr} = 1$. 
 Contours enclose 50\%, 70\%, and 99\% cumulative ranges of the parameter distribution, 
 while points denote systems lying outside the corresponding 99\% contour. 
 Black markers indicate GW190521 \citep{abbottGW190521BinaryBlack2020} and GW231123 \citep{abacGW231123BinaryBlack2025} events, 
 with error bars showing their observational uncertainties. 
 }
 \label{fig:samples_mockBBHfromAGN}
\end{figure}

More recently, \citet{vaccaroAGNdrivenBBHMergers2026} developed a more comprehensive semi-analytical framework that simultaneously accounts for BH capture, migration, binary pairing, gas-driven hardening, binary--single encounters, and hierarchical mergers, and systematically explored the dependence of the resulting BBH population on the SMBH mass, Eddington ratio, and disk viscosity. 
As shown in Figure~\ref{fig:samples_mockBBHfromAGN}, 
the distributions of simulated BBH samples demonstrate that AGN disks can produce an extended high-mass tail beyond the pair-instability mass gap, 
increasingly unequal mass ratios for high primary masses, and a correlation between the primary mass 
and $\lvert \chi_{\rm eff} \rvert$. 
The resulting mock BBH populations reach regions of parameter space broadly consistent with both GW190521 and GW231123, 
though such events are relatively rare and lie near the tails of the predicted distributions. 
In the low-viscosity models with $\alpha=0.01$, GW190521-like and GW231123-like events occur at levels of 
approximately $10^{-3}$ and a few $10^{-4}$, respectively, in the $(m_1,m_2)$ parameter space. 
These results thus suggest that AGN disks may provide a comparatively natural environment in which 
several extreme BBH properties can arise simultaneously. 

% %%%----------------------------------------------------
\subsubsection{Hierarchical Bayesian inference for the AGN formation channel}

In statistical analyses, consistency between observed data and competing theoretical models can be quantified through the Bayesian evidence, while the Bayes factor is commonly used for model comparison. 
The Bayes factor comparing population models ${\rm A}$ and ${\rm B}$ is defined as
\begin{equation}
\mathcal{B}_{\rm AB} = \frac{P(\{d_i\}| {\rm Model~A})}{P(\{d_i\}| {\rm Model~B})},
\end{equation}
where $\{d_i\}$ denotes the observed dataset of BBH mergers, and $P(\{d_i\}| {\rm Model~A})$ and $P(\{d_i\}| {\rm Model~B})$ are the Bayesian evidences of population models ${\rm A}$ and ${\rm B}$, respectively, given the dataset $\{d_i\}$.

Conditioned on the observation of $N_{\rm det}$ detections, the likelihood for a population model A 
can be written schematically as \citep{abacGWTC40PopulationProperties2026,abacGWTC50PopulationProperties2026}
\begin{equation}
\mathcal{L}(\{d_i\}| \boldsymbol{\Lambda}_{\rm A}) \propto
\prod_{i=1}^{N_{\rm det}}
\frac{
\int \! \mathcal{L}(d_i|\boldsymbol{\theta})\,
\pi(\boldsymbol{\theta}|\boldsymbol{\Lambda}_{\rm A})\,{\rm d}\boldsymbol{\theta}}
{
\displaystyle\beta(\boldsymbol{\Lambda}_{\rm A})},
\end{equation}
where the selection factor is
\begin{equation}
\beta(\boldsymbol{\Lambda}_{\rm A})=
\int \! p_{\rm det}(\boldsymbol{\theta})\,
\pi(\boldsymbol{\theta}|\boldsymbol{\Lambda}_{\rm A})\,{\rm d}\boldsymbol{\theta}.
\end{equation}
Here, $\boldsymbol{\Lambda}_{\rm A}$ denotes the hyperparameters characterizing population model A, 
and $\boldsymbol{\theta}$ denotes the intrinsic parameters of an individual event. The quantity $\mathcal{L}(d_i|\boldsymbol{\theta})$ is the event-level parameter-estimation likelihood, while $\pi(\boldsymbol{\theta}|\boldsymbol{\Lambda}_{\rm A})$ is the conditional prior of the intrinsic parameters under the population model A.
The detection probability can be expressed as 

\begin{equation}
p_{\rm det}(\boldsymbol{\theta})=
\int_{x(d)>x_{\rm thre}}p(d|\boldsymbol{\theta})\,{\rm d}d,
\end{equation}
where $x(d)$ is the detection statistic and $x_{\rm thre}$ is the adopted detection threshold. The precise threshold depends on the event catalog and search pipeline; in practical population analyses it is commonly defined through a false-alarm rate or an astrophysical-probability criterion.

Finally, the Bayesian evidence for model A, marginalized over its parameters, is given by

\begin{equation}
P(\{d_i\}| {\rm Model~A}) \!=\!\! 
\int \!\! {\rm d} \boldsymbol{\Lambda}_{\rm A}  \, \pi(\boldsymbol{\Lambda}_{\rm A}) \, 
\mathcal{L} \left( \{d_i\}| \boldsymbol{\Lambda}_{\rm A} \right), 
\end{equation}
where $\pi(\boldsymbol{\Lambda}_{\rm A})$ is the prior probability density distribution 
on the population model parameter set $\boldsymbol{\Lambda}_{\rm A}$. 

To investigate whether a fraction of the BBH merger events detected by the LVK network may originate from the AGN channel, 
one of the population models considered in a model-comparison analysis must represent the BBH population predicted by AGN-disk formation. 
In general, AGN-channel BBH population models can be constructed in two ways. 
The first is to use a population of BBHs generated directly from Monte Carlo or other numerical simulations, 
such as the AGN-channel BBH samples produced in \citet{delfaveroProspectsFormationGW2311232025} and \citet{vaccaroAGNdrivenBBHMergers2026} discussed above. 
Such simulated catalogs retain the multi-dimensional correlations among the intrinsic parameters that emerge from the underlying physical processes. 
The second approach is to construct a phenomenological or analytical population model based on the principal properties identified in the simulated AGN-channel population. 
Such a model can characterize the distributions of component masses, mass ratio, and spin parameters, together with the correlations among these parameters, without explicitly reproducing the full dynamical evolution of BBHs in the AGN disk.

The model against which the AGN-channel population is compared can be chosen according to the scientific question of interest. 
For example, one may compare the AGN-channel model with predictions from isolated binary evolution, 
dynamical assembly and hierarchical mergers in dense stellar environments such as globular clusters or nuclear star clusters, 
or other proposed BBH formation channels \citep{liComparingHierarchicalBlack2023, liOriginChannelsHierarchical2025, gayathriReconstructingOriginBlack2025, bertiInferringBlackHole2026}. 
Alternatively, the AGN population model can be compared directly with the population inferred from the BBH mergers detected by the LVK network, allowing one to assess whether the observed population contains a statistically significant contribution from the AGN channel \citep{tagawaSpinEvolutionStellarmass2020, gayathriBlackHoleMergers2021, liResolvingStellarCollapseHierarchicalMerger2024, liAlignedHierarchicalBlack2026}. 
In the former case, the analysis is primarily a comparison between competing formation-channel hypotheses, whereas in the latter case the goal is to determine the consistency of the AGN-channel population with the observed BBH population and, where possible, constrain the fraction of detectable mergers contributed by AGN disks.

% %%%----------------------
\paragraph{Bayesian inferences for AGN channel model using simulated BBH samples}

Several studies have attempted to determine whether a subpopulation of the BBH mergers observed by LVK 
is preferentially associated with the AGN formation channel by comparing the observed distributions of 
mass and spin parameters with populations generated from AGN-disk models. 
One of the early systematic studies was carried out by \citet{tagawaSpinEvolutionStellarmass2020}, 
who extended a semi-analytical model of BBH formation in AGN disks to follow 
the evolution of both the binary orbital angular momentum and the component spins. 
They found that the predicted effective-spin distribution can be broadly consistent 
with the BBH population observed in the early LIGO/Virgo runs. 
The study also identified potentially useful population-level signatures of the AGN channel, 
including relatively massive BBHs and correlations between mass and the width of the $\chi_{\rm eff}$ distribution. 
In addition, hierarchical mergers in AGN disks naturally produced systems with properties 
similar to GW190412 \citep{abbottGW190412ObservationBinaryBlackHole2020}, including an unequal mass ratio, 
a rapidly spinning primary, and a substantial in-plane spin component. 
These results suggested that the observed spin distribution, particularly when considered jointly with 
mass and mass ratio, could contain information about an AGN contribution to the BBH population. 

A more direct population-level analysis was performed by \citet{gayathriBlackHoleMergers2021}, 
who compared AGN-disk BBH populations with the BBH mergers observed in the LIGO/Virgo O1--O3a data. 
Using the predicted joint mass--spin distribution of AGN-assisted mergers, they found that 
an AGN-disk origin model was preferred over a phenomenological mass--spin model for 
approximately $20\%$ of the detected events, with a Bayes factor $\mathcal{B}>10$. 
They inferred an AGN BBH merger rate of $2.8 \pm 1.8\,{\rm Gpc}^{-3}\,{\rm yr}^{-1}$ and 
found that the AGN channel could naturally account for a substantial fraction of the systems 
with primary masses in the pair-instability mass gap. 
Although this analysis provided one of the first quantitative indications of an AGN-like 
contribution to the observed BBH population, it treated the AGN channel as a specific population model and 
therefore did not fully capture the theoretical uncertainty associated with the poorly constrained AGN-disk physics. 
A subsequent analysis by \citet{gayathriReconstructingOriginBlack2025} adopted a more flexible approach in which 
discrete population distributions generated from different astrophysical models 
were combined as fractional contributions to the observed BBH population. 
Using $87$ BBH detections from the O1--O3 observing runs, they jointly considered an AGN-assisted BBH population 
together with several isolated-binary populations generated with different SEVN assumptions 
for the common-envelope efficiency and metallicity. 
The inferred total BBH merger rate was $46.2\,{\rm Gpc}^{-3}\,{\rm yr}^{-1}$, 
with the AGN subpopulation contributing $21.2\,{\rm Gpc}^{-3}\,{\rm yr}^{-1}$ and 
the SEVN subpopulation contributing $25.0\,{\rm Gpc}^{-3}\,{\rm yr}^{-1}$. 
Thus, when theoretical model uncertainty and multiple formation channels are treated simultaneously, 
the observed BBH population remains consistent with a substantial AGN contribution, 
while the data do not require the entire population to originate from a single formation channel.

Subsequently, a more systematic hierarchical-Bayesian analysis was presented by \citet{liOriginChannelsHierarchical2025}, 
who inferred the relative contributions of AGN disks and nuclear star clusters to the hierarchical BBH population observed in the LVK O1--O3 runs. 
In their fiducial model, nuclear star clusters were inferred to dominate the intrinsic hierarchical-merger rate, 
with a contribution of $f_{\rm NSC} = 0.87^{+0.10}_{-0.29}$, while the AGN channel could contribute as much as 
$f_{\rm det,AGN} = 0.34^{+0.38}_{-0.26}$ of the detectable hierarchical mergers. 
They further found that hierarchical mergers themselves may account for at least $\sim 10\%$ of the detected BBH events in O1--O3. 
Importantly, however, their analysis also demonstrated that distinguishing the host environment solely from 
the observed BBH masses, mass ratios, and spins is challenging because different formation channels 
can produce substantially overlapping parameter distributions. 
Thus, this work provides evidence that an appreciable AGN contribution is allowed---and can be relatively large 
among detectable hierarchical mergers---but does not support a unique environmental identification of individual events.

More recently, \citet{heExploringHierarchicalMerger2026} investigated the hierarchical-merger interpretation of the asymmetric BBH events GW241011 and GW241110, 
which are characterized by rapidly spinning primaries, unequal component masses, and nonzero spin--orbit tilts. 
Using Bayesian model comparison, they found that both events favor a $2{\rm G}+1{\rm G}$ hierarchical-merger interpretation 
over a $1{\rm G}+1{\rm G}$ population, with $\ln\mathcal{B}^{2{\rm G}+1{\rm G}}_{1{\rm G}+1{\rm G}} \simeq  6.5-8.6$ 
for GW241011 and $\ln\mathcal{B}^{2{\rm G}+1{\rm G}}_{1{\rm G}+1{\rm G}} \simeq 3.0-4.5$ for GW241110, 
depending on the waveform model and environmental assumptions. 
When comparing $2{\rm G}+1{\rm G}$ populations generated for AGN disks and star clusters, 
the AGN models yielded slightly larger evidence, mainly because of their predicted spin-tilt distributions. 
%Nevertheless, the data do not provide decisive evidence for an AGN origin, demonstrating again that evidence 
%for hierarchical assembly does not automatically translate into evidence for a specific astrophysical environment.

Taken together, these studies provide a progressively stronger, but still non-unique, 
case for an AGN contribution to the observed BBH population. 
The evidence is particularly suggestive for systems occupying regions of parameter space associated with hierarchical growth. 
Nevertheless, the inferred AGN fraction depends strongly on the adopted AGN and competing population models, 
on assumptions about hierarchical-merger efficiencies, and on the degree of overlap between different formation channels. 
Therefore, the current observational picture is best summarized as evidence that 
a non-negligible subset of LVK BBHs may be compatible with an AGN-disk origin, 
rather than as evidence that any particular set of events has been uniquely identified as AGN-produced. 

% %%%----------------------
\paragraph{Bayesian inferences for AGN channel model via phenomenological analytic formulations}

A complementary approach to simulation-based AGN population catalogs is to encode 
the principal observational signatures of the AGN channel directly into phenomenological population models 
or parameterized correlations among BBH intrinsic parameters. 
This approach is particularly useful when the detailed AGN-disk dynamics remain uncertain, 
because it allows one to test whether the characteristic correlations expected from hierarchical growth 
and gas-assisted spin evolution are already present in the observed LVK population. 
Several recent studies have found evidence for subpopulations with properties that are compatible 
with AGN-assisted hierarchical mergers, although the statistical strength of the evidence and 
the degree to which the AGN interpretation is unique vary substantially among different analyses. 

\citet{liResolvingStellarCollapseHierarchicalMerger2024} introduced a flexible semi-parametric population model to 
investigate the joint mass and spin properties of BBHs in GWTC-3. 
Their analysis strongly favored the existence of two distinct BBH subpopulations over 
a single-population model, with $\ln\mathcal{B}=7.5$. 
The higher-mass subpopulation spans approximately $25-80\,M_\odot$ and is characterized by 
substantially larger spin magnitudes, with a characteristic value of $\chi\sim0.75$, 
in qualitative agreement with expectations for hierarchical mergers. 
The result therefore provides population-level evidence for a high-mass, high-spin subpopulation that 
is difficult to associate with purely first-generation BBHs. 
Because hierarchical mergers can occur in both dense stellar systems and AGN disks, 
however, this result should primarily be regarded as evidence for a hierarchical component 
rather than as a unique identification of an AGN origin. 

\citet{liLiRevealingHeffCorrelation2025} subsequently revisited the previously reported $\chi_{\rm eff}-q$ correlation 
by allowing different effective-spin distributions for different mass subpopulations. 
They found that the apparent global correlation between $\chi_{\rm eff}$ and mass ratio becomes substantially weaker, 
and may disappear, once a second spin population associated with the high-mass BBHs is introduced. 
The low-mass component has a narrow $\chi_{\rm eff}$ distribution centered near $0.05$ and 
a primary-mass cutoff around $\sim 40\,M_\odot$, 
whereas the high-mass component has a much broader effective inspiral spin distribution with a positive peak at 
$\mu_{\chi} \sim 0.4$. 
The positive mean of the high-mass component is favored at the $98\%$ credibility level, 
and a symmetric spin distribution is disfavored with a Bayes factor of $\ln\mathcal{B}=1.5$. 
The authors argue that this high-mass, broad, positively shifted spin component is consistent with hierarchical mergers in AGN disks, 
while the presence of negative-$\chi_{\rm eff}$ systems indicates that hierarchical mergers 
in dynamical environments such as star clusters may also contribute. 
Thus, the key implication is that a phenomenological decomposition of the observed population can reveal 
an AGN-compatible hierarchical component even without explicitly modeling the AGN disk dynamics. 

A more direct test of the AGN interpretation was presented by \citet{liAlignedHierarchicalBlack2026}, 
who analyzed GWTC-4.0 with a flexible mixture model for component masses, spin magnitudes, and spin tilt angles. 
They identified a second subpopulation characterized by high spin magnitudes, $\chi\sim0.8$, 
and a broad primary-mass distribution extending to $\gtrsim150\,M_\odot$. 
Within this subpopulation, the data show a pronounced preference for spins aligned with the orbital angular momentum: 
an isotropic tilt distribution is disfavored with $\ln\mathcal{B}=4.5$. 
The aligned systems constitute approximately $0.57^{+0.23}_{-0.31}$ of the high-spin subpopulation, 
corresponding to a local merger rate of about $0.25^{+0.38}_{-0.16}\,{\rm Gpc}^{-3}\,{\rm yr}^{-1}$. 
The combination of high masses, high spins, and preferential spin alignment is naturally expected for 
hierarchical mergers embedded in AGN disks, where gas torques can align BH spins with the disk angular momentum. 
This study therefore provides one of the more direct population-level indications in favor of an AGN-disk contribution, 
although the inferred subpopulation itself is not uniquely identified with AGN disks and could in principle contain contributions from other hierarchical channels.

A complementary approach was adopted by \citet{bertiInferringBlackHole2026}, who parameterized the correlations 
between BBH spin magnitudes and masses, together with the distributions of spin orientations, 
according to the expectations of different formation scenarios, and applied hierarchical Bayesian inference to GWTC-4.0. 
They found strong evidence that the BBH population contains a positive correlation between mass and spin magnitude, 
and that a hierarchical-merger scenario provides a better fit to the observations 
because successive merger generations naturally produce higher spins at larger masses. 
However, the current data are not sufficiently informative to distinguish the AGN scenario, 
characterized by preferentially aligned spins, from the approximately isotropic-spin hierarchical-merger scenario 
expected in dense stellar environments, as the two models yield comparable Bayesian evidence. 
A related result was obtained by \citet{roupasEvidenceAccretiondrivenSubpopulation2026}, who tested a theoretically motivated 
accretion-driven spin--mass relation against the GWTC-5.0 catalog. In their model, black holes initially 
have low spins at low masses, while accretion progressively increases the spin magnitude with increasing mass, 
eventually approaching a high-spin saturation regime. The analysis found evidence for an accretion-driven subpopulation, 
with a preferred transition mass of $m^{t}=20.7^{+11.6}_{-1.2}\,M_\odot$ at $90\%$ credibility 
and a Bayes factor of $\ln\mathcal{B}=5.5$ for the presence of a transition in the population. 
Above this mass scale, the evidence for an accretion-driven component becomes particularly strong, 
reaching $\ln\mathcal{B}=16.6$ relative to the LVK fiducial spin population. 
Ten events show positive support for the accretion-driven interpretation, 
with weaker evidence extending across a broader range of component masses. 
Although the accretion model considered in \citet{roupasEvidenceAccretiondrivenSubpopulation2026} is based on gas-assisted growth 
in the birth environments of star clusters rather than AGN disks specifically, 
its predicted mass-dependent spin-up provides an observationally relevant phenomenology that 
is also naturally expected in AGN environments. Taken together, these studies strengthen the evidence that 
the observed BBH population contains a nontrivial correlation between mass and spin magnitude and 
that gas-assisted and/or hierarchical growth may contribute to this structure. 
At the same time, the results do not uniquely identify AGN disks as the origin of 
the inferred high-spin or accretion-driven subpopulation, 
since similar mass--spin correlations can arise from other astrophysical environments.

\citet{bartosHighSpinBBHSubpopulation2026} considered a somewhat different AGN signature: 
spin-up through gas accretion rather than exclusively spin growth through hierarchical mergers. 
Using a three-component mixture model for 166 LVK BBH mergers, they fixed the component shapes to 
theoretically motivated spin-magnitude distributions and inferred only the mixing fractions. 
They found evidence for a distinct high-spin subpopulation with $\ln\mathcal{B}=5.7$, 
comprising approximately $10\%$ of the detected BBH mergers, with a $90\%$ credible interval of about $1\%-14\%$. 
The spins of this component cluster near $a_1\simeq0.9$, while the lower-spin hierarchical-merger expectation of 
$a_1\simeq0.7$ is strongly disfavored as the location of this particular high-spin component. 
The candidate events also have systematically higher primary masses, with a median $m_1 \simeq 58\,M_\odot$, 
and more positive effective spins, with a median $\chi_{\rm eff}\simeq0.33$, than the standard-dominated population. 
The result is especially interesting for the AGN channel because efficient gas accretion can 
simultaneously increase the BH mass and spin and align the spin with the disk angular momentum. 
Notably, the candidate accretion population is not restricted to systems above the pair-instability mass gap, 
and GW190521 receives comparable support for an accretion interpretation. 
This provides an independent phenomenological route to an AGN-related subpopulation, 
complementary to the hierarchical-merger interpretation. 

\citet{rayAstrophysicalOriginBinary2026} and \citet{qiuchengReversiblejumpMCMCReveals2026} independently investigated 
the substructure of the BBH population using flexible, data-driven mixture models, and 
both identified a subdominant component with properties relevant to hierarchical formation 
and therefore potentially compatible with an AGN contribution. 
\citet{rayAstrophysicalOriginBinary2026} identified three astrophysically distinct subpopulations in the LVK catalog, 
with the lowest-mass population around $m_1\sim10\,M_\odot$ broadly consistent with isolated binary evolution, 
a component around $m_1\sim35\,M_\odot$ associated primarily with dynamical formation in globular clusters, 
and a higher-generation component contributing preferentially at larger masses. 
These subpopulations exhibit distinct distributions of mass ratio, spin alignment, spin precession, and redshift evolution, with inferred underlying fractions of approximately $79.0\%, 14.5\%,$ and $2.5\%$, respectively. 
\citet{qiuchengReversiblejumpMCMCReveals2026}, using a minimally parametrized reversible-jump Markov chain Monte Carlo framework, 
likewise found evidence for multiple BBH subpopulations, including a low-mass population with preferential spin alignment, 
an intermediate-mass population with approximately isotropic spins and a preference for equal masses, 
and a broad high-spin population extending over a wide range of primary masses. 
The latter component is particularly relevant to AGN-disk scenarios 
because both hierarchical mergers and gas-assisted evolution can produce high-spin BHs, 
while the mass, mass-ratio, and spin properties of the high-generation component 
overlap substantially with the predictions of AGN-assisted hierarchical growth. 
Neither study, however, uniquely identifies this subpopulation with AGN disks; 
rather, their results provide independent, data-driven evidence for a subdominant 
or hierarchical component in the LVK BBH population whose properties are consistent with, 
among other possibilities, an AGN contribution. 

\begin{figure}[htbp] 
 \centering
 \includegraphics[width=0.7\textwidth]{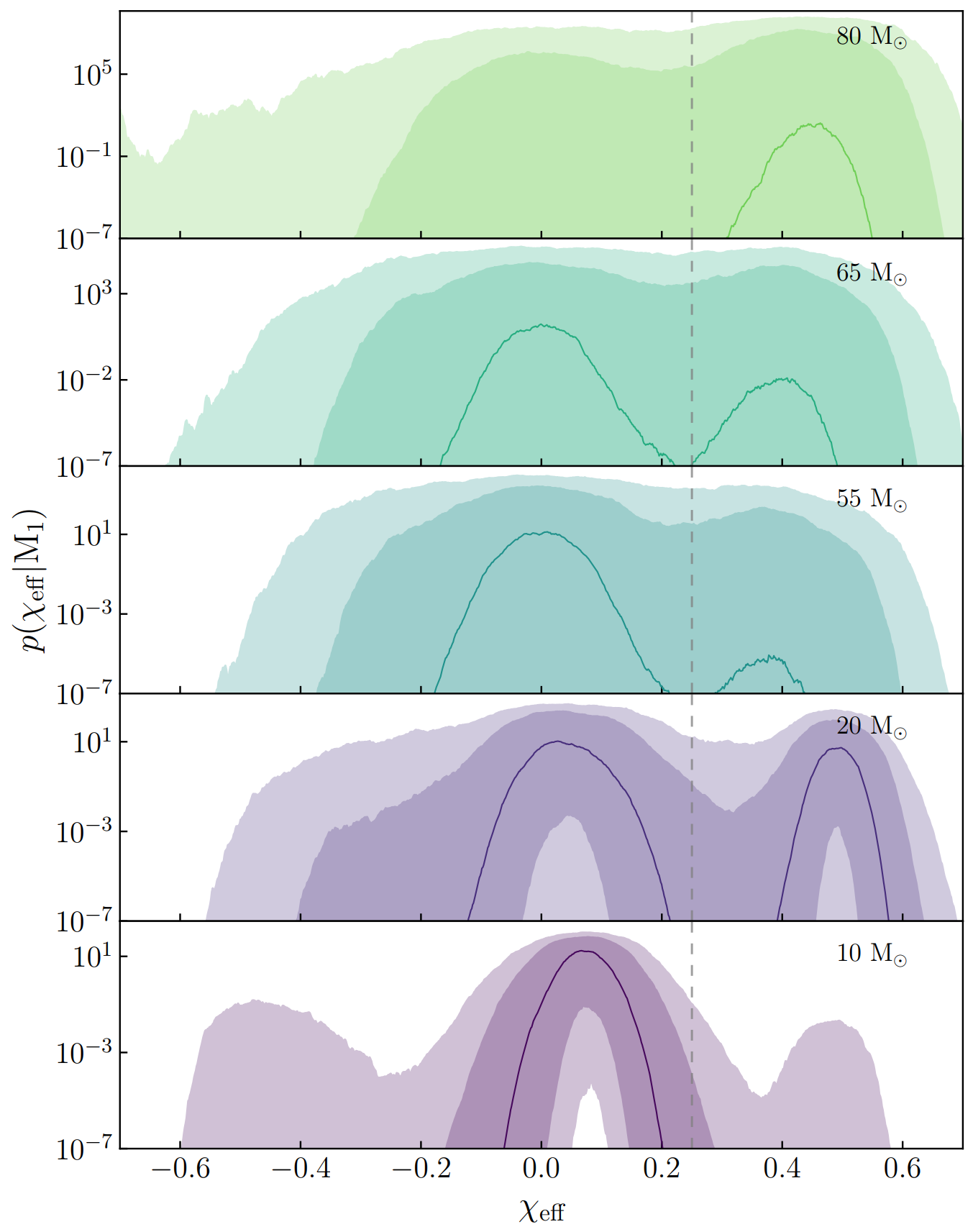}
 \caption{Reconstructed $\chi_{\rm eff}$ distribution conditioned on different primary masses of BBH events \citep{rinaldiWhenBlackHoles2026}. 
 The vertical dashed line at $\chi_{\rm eff} = 0.25$ marks the referenced boundary between the two subpopulations. 
 }
 \label{fig:m1chieff_relation}
\end{figure}

Using the larger GWTC-5.0 catalog, \citet{alvarez-lopezEvidenceAdditionalStructure2026} further investigated the structure 
of the effective-spin distribution and showed that its apparent skewness is more naturally interpreted 
as additional structure superimposed on a nearly symmetric, low-$\chi_{\rm eff}$ Gaussian bulk. 
Their flexible and parametric analyses both found tentative evidence for a mass-dependent excess of positive over negative $\chi_{\rm eff}$ outside the Gaussian component. 
In particular, the mass range $m_1\simeq46-65\,M_\odot$ requires a negative-$\chi_{\rm eff}$ component at odds of approximately $23\!:\!1$. 
A complementary data-driven analysis by \citet{rinaldiWhenBlackHoles2026} further examined 
the joint mass--effective-spin distribution with minimal assumptions on its functional form and 
found evidence for multiple BBH subpopulations with distinct spin properties. 
As shown in Figure~\ref{fig:m1chieff_relation}, 
they found that a high-mass component exhibits a preference for positive $\chi_{\rm eff}$, suggesting that
the spin distribution of massive BBHs may retain a preferred angular-momentum direction rather than being fully isotropic. 
If the non-Gaussian high-$|\chi_{\rm eff}|$ component is associated with hierarchical mergers, 
the excess of positively aligned spins at higher masses is qualitatively consistent with environments that 
can imprint a preferred angular-momentum axis through gas interactions, such as AGN disks. 
These results therefore provide an important, largely model-independent connection between the spin substructure 
inferred directly from GWTC-5.0 and the spin signatures expected from AGN-assisted hierarchical mergers, 
while remaining insufficient to establish an AGN origin on their own. 

An important caveat is provided by \citet{wolfeNoModelindependentEvidence2026}, who re-examined claims of 
a mass-dependent peak in the spin-tilt distribution using GWTC-4.0. 
They found that a preferred peak in spin misalignments is not statistically significant 
in a model-independent analysis and found no confident mass--tilt correlation, 
although the positive correlation between spin magnitude and mass remains supported. 
This result cautions against treating apparent spin-alignment substructure as 
a robust discriminator of the AGN channel with current data, and 
motivates relying on the joint mass--spin distribution rather than on spin tilts alone.

A complementary, more model-focused analysis was presented by \citet{plunkettSignaturesSubpopulationHierarchical2026}, 
who modeled the joint distribution of the effective inspiral spin $\chi_{\rm eff}$ and precession spin $\chi_p$ 
in GWTC-4.0 using an astrophysically motivated prescription for hierarchical mergers, 
with $\chi_p$ defined as
\[ \chi_p \equiv \max \left[\chi_1 \sin\theta_1, \frac{q(4q+3)}{4+3q} \chi_2 \sin\theta_2 \right]. \]
Their analysis found decisive evidence for a transition in the BBH population at $m_1 \sim 46.2 \,M_\odot$, 
above which the population is inferred to be dominated by higher-generation mergers. 
They also identified a peak in the hierarchical-merger rate at $m_1 \sim 15.7 \,M_\odot$. 
The inferred low- and high-mass subpopulations of hierarchical mergers suggest 
contributions from stellar systems with different metallicities. 
Although this result does not uniquely identify the hierarchical subpopulation with AGN disks, 
it provides independent evidence that a substantial component of the LVK BBH population 
has properties characteristic of hierarchical assembly. 
Such a component overlaps naturally with the parameter space predicted 
for AGN-assisted hierarchical mergers, and therefore strengthens the motivation 
for considering AGN disks as one possible astrophysical environment for this subpopulation. 

Taken together, phenomenological population studies provide a qualitatively consistent picture in which 
the LVK BBH catalog contains multiple subpopulations with distinct mass and spin properties. 
Several analyses identify a high-mass, high-spin component associated with higher-generation mergers, 
as well as a population transition above $\sim 45\,M_\odot$ consistent with the onset of hierarchical mergers. 
The observed subpopulations exhibit combinations of high primary mass, enhanced spin magnitude, 
positive $\chi_{\rm eff}$, spin alignment, and, in some analyses, 
unequal mass ratios and broad precession-spin distributions. 
These properties are naturally produced when BBH growth proceeds through successive mergers and, 
in an AGN environment, can be further modified by gas accretion and spin alignment. 
The evidence is therefore strongest when several observables are considered jointly rather than individually. 
At the same time, the present analyses do not imply that the relevant high-mass 
or high-spin systems are uniquely AGN-produced: 
hierarchical mergers in stellar clusters and other dense environments can generate many of the same mass and spin signatures. 
Consequently, these phenomenological and data-driven analyses provide evidence for 
a hierarchical or AGN-compatible subpopulation in the LVK BBH catalog, 
but the available data are not yet sufficient to identify AGN disks as one of its definitive astrophysical origins.

%%%====================================================
\subsection{Statistical constraints for the BBH-AGN spatial correlation}     \label{sec:sta_corr}

The complex astrophysical processes governing BBH formation and evolution in AGN disks 
introduce substantial uncertainties into the population properties predicted by AGN-channel models. 
These uncertainties have motivated the development of tests of the connection between BBHs and AGNs that 
are less dependent on the detailed assumptions of any particular BBH population model. 
One such model-independent approach is based on spatial correlations between BBH merger events and AGN populations. 
The underlying premise is straightforward: if a detected BBH merger is genuinely formed in an AGN environment, 
the BBH and its host AGN must occupy the same physical location. 
In principle, identifying a statistical spatial correlation between the sky positions and 
distances of BBH mergers and those of AGNs can therefore provide an independent test of the AGN formation channel. 

In practice, however, the relatively poor localization accuracy of the current LVK detector network poses a major limitation. 
The three-dimensional localization regions of individual BBH mergers are typically sufficiently 
large to contain roughly $10^2-10^5$ AGNs, making it generally impossible to unambiguously identify 
a unique AGN host for an individual BBH event. 
Consequently, spatial-correlation studies cannot presently establish one-to-one associations 
between individual BBH mergers and their putative AGN hosts. 
Instead, they must rely on statistical population-level analyses, 
testing whether the observed spatial distribution of BBH mergers exhibits an excess correlation with 
the distribution of AGNs beyond that expected from chance alignments and the underlying large-scale structure. 
This approach is therefore complementary to population-model-based inference and, 
with improved GW localization from future detector networks, 
may eventually provide a more direct and comparatively model-independent probe 
of the AGN contribution to the BBH merger population. 

The statistical framework for testing spatial correlations between BBH mergers and AGNs 
was originally developed in analogy with spatial-correlation searches for associations 
between astrophysical neutrinos and their potential counterparts or source populations 
\citep{bartosGravitationalwaveLocalizationAlone2017, palmeseLIGOVirgoBlack2021}. 
The motivation is similar in both cases: the relatively poor angular localization of high-energy neutrinos 
and GW sources generally prevents a unique association with an individual astrophysical source, 
but a statistical excess of spatially coincident events can nevertheless reveal a population-level connection. 
The first dedicated statistical framework for testing a BBH--AGN spatial correlation 
was proposed by \citet{bartosGravitationalwaveLocalizationAlone2017}, who demonstrated that the spatial distribution of 
GW localization regions and AGNs can be used to statistically constrain the fraction of BBH mergers 
occurring in AGN hosts, even when individual GW events cannot be uniquely associated with specific AGNs. 
Subsequently, \citet{palmeseLIGOVirgoBlack2021} incorporated the BBH--AGN association problem into a Bayesian inference framework. 
Although the formulations of \citet{bartosGravitationalwaveLocalizationAlone2017} and \citet{palmeseLIGOVirgoBlack2021} 
differ in their statistical implementation, their central idea is the same. 

%%%----------------------------------------------------
\subsubsection{Statistical framework for testing the BBH-AGN spatial correlation}

Here, we select the framework of \citet{bartosGravitationalwaveLocalizationAlone2017} as a representative example 
to introduce the basic principle and practical implementation of the BBH-AGN spatial-correlation analysis. 
The basic idea of the spatial-correlation test is to compare the number density of AGNs around 
the localization volume of each GW event with that expected from chance coincidence. 
For the $i$-th GW event, let $\Delta V_i$ denote its three-dimensional localization comoving volume 
and let $N_{{\rm AGN},i}$ be the number of AGNs contained within this volume. 
\citet{bartosGravitationalwaveLocalizationAlone2017} assume a uniform AGN number density $\rho_{\rm AGN}$. 
Under the null hypothesis that the GW event is not produced in an AGN, 
the number of AGNs in the localization volume follows a Poisson distribution with mean $\rho_{\rm AGN} \Delta V_i$, 
giving the background probability 
\begin{equation}   \label{eq:poss_bi}
B_i(N_{{\rm AGN},i}) = {\rm Poiss} 
\left( N_{{\rm AGN},i}, \rho_{\rm AGN} \Delta V_i  \right).
\end{equation}

Conversely, if the GW event originates in an AGN, one AGN is guaranteed to reside at the true source position, 
while the remaining AGNs are distributed according to the same Poisson background. 
The corresponding signal probability is therefore 
\begin{equation}   \label{eq:poss_si}
S_i(N_{{\rm AGN},i}) = {\rm Poiss} 
\left( N_{{\rm AGN},i}-1, \rho_{\rm AGN} \Delta V_i \right).
\end{equation}
Thus, for a genuine GW--AGN association, the observed AGN count is statistically shifted upward relative to the background expectation.

For a population of $N_{\rm GW}$ detected BBH mergers, the formation channel of an individual event cannot in general be classified as AGN or non-AGN, because of the large localization volumes. 
Instead, one introduces $f_{\rm AGN}$, the fraction of BBH mergers assumed to originate in AGNs. 
The likelihood for the complete GW sample is then 
\begin{equation}
\mathcal{L}(f_{\rm AGN}) = \prod_{i=1}^{N_{\rm GW}}
\left[ f_{\rm AGN}S_i + (1-f_{\rm AGN}) B_i \right].
\end{equation}

The subsequent studies \citep{veronesiMostLuminousAGN2023, veronesiConstrainingAGNFormation2025} modified the quantification of 
the signal probability by treating the AGNs within the GW localization volume as sampling points and 
performing a discrete integral over the three-dimensional localization probability density of the BBH merger. 
The corresponding background probability was defined as the expected value of this integral, 
namely the confidence level of the GW localization volume being integrated over. 
This formulation has the advantage of making more direct use of the positional information encoded in 
the GW localization probability distribution. In particular, an AGN located in a region of 
higher GW localization probability density contributes more highly to the signal statistic $S_i$, 
thereby assigning greater weight to AGNs that are more likely to coincide with the true BBH merger location.
However, in both \citet{veronesiMostLuminousAGN2023} and \citet{veronesiConstrainingAGNFormation2025}, the calculation of $S_i$ still 
relies on the simplifying assumption that AGNs are uniformly distributed throughout the Universe. 
This assumption can differ substantially from the spatial distribution of AGNs in real observational catalogs. 
In practice, the sensitivity limits of astronomical surveys introduce significant selection effects, 
so the AGNs recorded in a survey catalog generally exhibit a spatial distribution that is neither homogeneous nor isotropic. 
The observed AGN density can vary systematically with sky position and redshift. 
Consequently, directly applying a uniform AGN-density assumption to an observational catalog 
may cause the probability of chance spatial coincidences to be misestimated and 
may therefore bias the inferred significance of a BBH--AGN spatial correlation. 

To mitigate this issue, \citet{zhuEvidenceFractionLIGO2025} retain the same signal--background structure 
but generalize it to an inhomogeneous, observationally incomplete AGN catalog. 
Instead of describing the signal simply by the total number of AGNs in the localization volume, 
they assign a local AGN number density $n_{\rm AGN}(\mathbf{x})$ to each cataloged AGN position $\mathbf{x}_j$, 
estimated using a three-dimensional Voronoi tessellation. 
Their signal statistic is therefore constructed as a sum over the AGNs inside the localization volume, 
\begin{equation}   \label{eq:int_si}
S_i = \sum_{j=1}^{N_{{\rm AGN},i}} \!\!
\frac{p_i(\mathbf{x}_j)}{n_{\rm AGN}(\mathbf{x}_j)},
\end{equation}
where $p_i(\mathbf{x})$ is the three-dimensional localization probability density of the $i$-th GW event. 
The corresponding background statistic is
\begin{equation}   \label{eq:int_bi}
B_i = 0.9 f_{{\rm cover},i}, 
\end{equation}
where the factor $0.9$ corresponds to the adopted $90\%$ GW localization credible volume, and 
$f_{{\rm cover},i}$ is the fraction of the $i$-th BBH's localization error volume covered by the surveyed AGN catalog. 

The approaches represented by Eqs. (\ref{eq:poss_bi}, \ref{eq:poss_si}) and (\ref{eq:int_si}, \ref{eq:int_bi}) differ mainly in the level of realism with which the AGN spatial distribution is modeled. 
The former treat AGNs as a population with an approximately uniform number density 
and use the total AGN count within each GW localization volume as the primary statistic. 
The latter, in contrast, use the actual three-dimensional distribution of cataloged AGNs and 
the full GW localization probability distribution to construct a spatially weighted signal statistic, 
allowing the method to account for strong spatial variations in AGN density and incomplete catalog coverage. 
In both cases, however, the essential test is identical: if a non-negligible fraction of BBH mergers originates in AGNs, the observed GW localization regions should contain a statistically significant excess of AGNs relative to 
randomly placed localization regions. 

%%%----------------------------------------------------
\subsubsection{Preliminary evidence for a spatial correlation between BBHs and AGNs}

The possibility of identifying the BBH formation environment through 
spatial correlations with AGNs was first explored in a forecasting context, 
before sufficiently large GW catalogs were available for a direct observational test. 
\citet{corleyLocalizationBinaryBlack2019} showed that the localization of BBH mergers can be improved 
when additional information, such as an independently constrained binary inclination, 
is available, thereby strengthening the prospects for associating GW events with their host environments. 
Building on the statistical-correlation framework introduced by \citet{bartosGravitationalwaveLocalizationAlone2017}, 
\citet{veronesiDetectabilitySpatialCorrelation2022} performed a dedicated detectability study for GW--AGN spatial correlations 
using simulated BBH catalogs and simulated detector localizations representative of the O3 and O4 observing networks. 
They found that the ability to detect a spatial correlation depends sensitively on the AGN number density, 
the AGN luminosity threshold, the fraction of BBH mergers originating in AGNs, and the GW localization volume. 
For the expected O3 catalog, the simulated number of BBH detections was insufficient to exclude 
the null hypothesis for relatively common, moderately luminous AGNs at high significance, 
whereas substantially rarer and more luminous AGN populations could already be tested with a modest number of detections. 
The study therefore established the observational feasibility of spatial-correlation tests and 
anticipated that the increasing number of GW events and improved localization in subsequent 
observing runs would make direct constraints on an AGN contribution possible. 

The first direct applications to observed BBH events did not find evidence for a significant spatial association, 
but instead placed increasingly stringent upper limits on the contribution of AGNs. 
\citet{veronesiMostLuminousAGN2023} cross-matched 30 low-redshift BBH mergers from O1--O3 with 
all-sky catalogs of spectroscopically confirmed, high-luminosity AGNs. 
They obtained upper limits of $f_{\rm AGN}<0.49$ for AGNs with bolometric luminosity $L_{\rm bol}>10^{45.5}\,{\rm erg\,s^{-1}}$ and 
$f_{\rm AGN}<0.17$ for the even more luminous population with $L_{\rm bol}>10^{46}\,{\rm erg\,s^{-1}}$ at 95\% credibility. 
These results excluded the possibility that the brightest local AGNs host the majority of the detected BBH mergers, 
although they did not rule out a contribution from lower-luminosity AGNs. 
\citet{veronesiConstrainingAGNFormation2025} subsequently extended the analysis to 159 GW mergers and 
the much larger and more homogeneous Quaia quasar catalog, reaching $z\lesssim1.5$. 
With the improved redshift coverage and catalog completeness, they obtained even tighter upper limits: 
unobscured AGNs with $L_{\rm bol}>10^{44.5}\,{\rm erg\,s^{-1}}$ and $L_{\rm bol}>10^{45}\,{\rm erg\,s^{-1}}$ 
contribute no more than $21\%$ and $11\%$, respectively, of the detected GW population at 95\% credibility. 
Together, these studies showed that the most luminous AGNs cannot account for a dominant fraction of 
the observed BBH mergers, but they left open---and indeed could not directly test---the possibility that 
BBHs preferentially form in the much more numerous lower-luminosity or lower-accretion-rate AGNs. 

\begin{figure}[htbp] 
 \centering
 \includegraphics[width=0.7\textwidth]{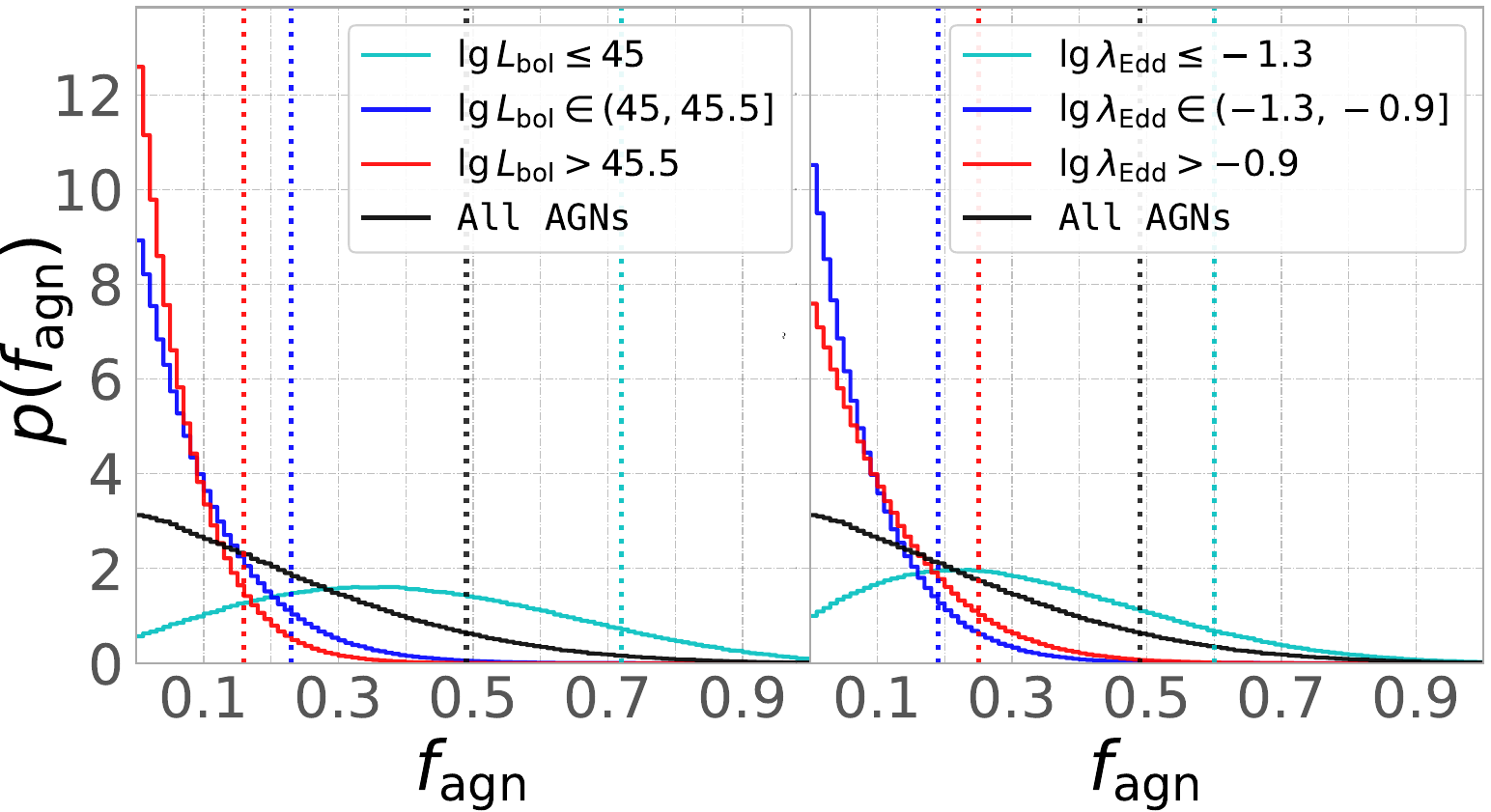}
 \caption{Probability density distributions of $f_{\rm AGN}$ inferred from \citep{zhuEvidenceFractionLIGO2025}. 
 Different curves correspond to lower (cyan), moderate (blue), and higher (red) $L_{\rm bol}$ (left panel) 
 or $\lambda_{\rm Edd}$ (right) AGN sub-catalogs. 
 Results for the full catalog are shown with black curves. 
 Vertical dotted lines mark the 90\% confidence upper limits. 
 }
 \label{fig:fagn_infer}
\end{figure}

A qualitatively different result was reported by \citet{zhuEvidenceFractionLIGO2025}, 
who analyzed the spatial correlation between GW localizations from O1--O4a and the SDSS DR16 AGN catalog. 
Rather than restricting the analysis to the rare, brightest quasars, 
they considered substantially lower-luminosity and lower-Eddington-ratio AGNs and 
found preliminary evidence for an excess of such AGNs within the localization regions of the observed GW events. 
As shown in Figure~\ref{fig:fagn_infer}, 
for AGNs with bolometric luminosity $10^{44.5}\lesssim L_{\rm bol}\lesssim 10^{45}\,{\rm erg\,s^{-1}}$, the observed excess 
is explained by an AGN-origin fraction of $f_{\rm AGN}=0.39^{+0.41}_{-0.32}$, 
while for AGNs with Eddington-ratio $0.01 \lesssim \lambda_{\rm Edd} \lesssim 0.05,$ 
the corresponding fraction is $f_{\rm AGN}=0.29^{+0.40}_{-0.25}$, both quoted at 90\% confidence. 
Here $\lambda_{\rm Edd} \equiv L_{\rm bol} / L_{\rm Edd}$, 
where $L_{\rm bol} = 1.3 \times 10^{38} (M_{\rm BH}/M_{\odot})~\!{\rm erg~s}^{-1}$ 
is the Eddington luminosity of a black hole with mass $M_{\rm BH}$. 
Monte Carlo realizations were used to assess the probability that the observed excess could arise from chance alignments, 
and the results favored a genuine spatial correlation over a purely random coincidence. 
These findings therefore provide the first reported evidence for a nonzero BBH--AGN spatial correlation, 
while still implying a substantial uncertainty in the inferred AGN fraction. 

The apparent difference between the results of \citet{veronesiMostLuminousAGN2023}, \citet{veronesiConstrainingAGNFormation2025} 
and \citet{zhuEvidenceFractionLIGO2025} is largely attributable to the AGN populations 
probed by the different analyses rather than to a direct contradiction. 
The former studies primarily constrained relatively luminous, unobscured AGNs and 
found that these systems cannot dominate the BBH merger rate; 
the latter explicitly extended the search into the lower-luminosity and lower-Eddington-ratio regime, 
where the AGN population is substantially more numerous and may be more representative of the environments 
relevant for BBH formation in accretion disks. 
In addition, \citet{zhuEvidenceFractionLIGO2025} used the full O1--O4a GW sample and the SDSS DR16 AGN catalog, 
and their spatial statistic makes use of the three-dimensional localization probability together with 
the local AGN density, rather than relying primarily on the total number of AGNs within a localization volume. 
Their result therefore has greater sensitivity to relatively weak but spatially coherent excesses of AGNs. 
The overall picture that emerges is consequently consistent: 
current spatial-correlation studies disfavour the hypothesis that 
the majority of BBH mergers originate in the most luminous AGNs, 
while leaving---and potentially supporting---a non-negligible contribution from 
lower-luminosity or lower-accretion-rate AGNs. 
The latter population may therefore provide a more promising observational target 
for testing the AGN formation channel in future, 
especially as GW localization improves and AGN catalogs become more complete. 

%%%----------------------------------------------------
\subsubsection{Preliminary indication for a spatial correlation between BBH mergers and AGN flares}

The possibility of using AGN flares as electromagnetic counterparts to BBH mergers provides 
a complementary way to probe the AGN formation channel. 
\citet{palmeseLIGOVirgoBlack2021} first developed a Bayesian framework to assess 
the statistical significance of associations between GW mergers and AGN flares, 
motivated in particular by the proposed association between GW190521 and an AGN flare. 
They emphasized that an apparent spatial and temporal coincidence does not by 
itself constitute evidence for a physical association, because the large three-dimensional 
localization volume of GW190521 contains thousands of potential AGN hosts. 
Indeed, the $90\%$ localization volume of GW190521 was estimated to contain approximately 7400 unobscured 
AGNs brighter than $g=20.5 \, {\rm mag}$, implying a substantial probability of chance coincidence. 
Their forecasting analysis showed that a confident association could nevertheless be established 
statistically with a sufficiently large sample of well-selected GW events and electromagnetic follow-up observations. 
In particular, if the fraction of BBH mergers producing observable AGN flares is larger than $\sim10\%$, 
targeted follow-up of a suitably selected subset of well-localized and relatively massive BBH mergers 
could potentially establish a significant association during subsequent observing runs. 
Thus, this work established the statistical basis and observational prospects for 
testing BBH--AGN flare correlations at the population level, 
while highlighting the importance of accounting for chance coincidences. 

The first analyses using actual GW and AGN-flare catalogs did not find 
statistically significant evidence for a GW--flare association. 
\citet{veronesiAGNFlaresCounterparts2024} analyzed 78 BBH mergers detected during O3 together with 
the 20 unusual AGN flares identified by \citet{grahamLightDarkSearching2023} in ZTF data. 
Their spatial and temporal correlation analysis yielded a posterior for 
the fraction $f_{\rm flare}$ of BBH mergers producing an observable AGN flare that peaked at zero, 
with a $90\%$ upper limit of $f_{\rm flare}<0.155$. 
They further showed that the fact that the candidate flares preferentially occurred within 
the localization regions of relatively massive BBHs did not by itself indicate a physical connection, 
since more massive GW systems tend to have larger localization volumes and 
therefore a higher probability of containing unrelated AGN flares. 
\citet{cabreraMultimessengerConstraintsLIGO2026} subsequently performed an independent population-level analysis 
using the GWTC-3 BBH catalog and the same ZTF AGN-flare sample. 
Their fiducial analysis also found a posterior maximized at zero, 
with a $90\%$ upper limit of $f_{\rm flare}<2.8\%$ for the fraction of LVK BBHs producing detectable AGN flares. 
Importantly, the absence of an observable flare does not imply that the corresponding BBH did not form in an AGN disk: 
accounting for the limited electromagnetic observability of merger-induced flares, 
their analysis still allowed up to $\sim40\%$ of BBH mergers to originate in AGN disks. 
Moreover, individual GW--flare coincidences were generally more likely to arise 
from the background AGN-flare population than from a causal BBH--AGN connection. 

\begin{figure}[htbp] 
 \centering
 \includegraphics[width=0.7\textwidth]{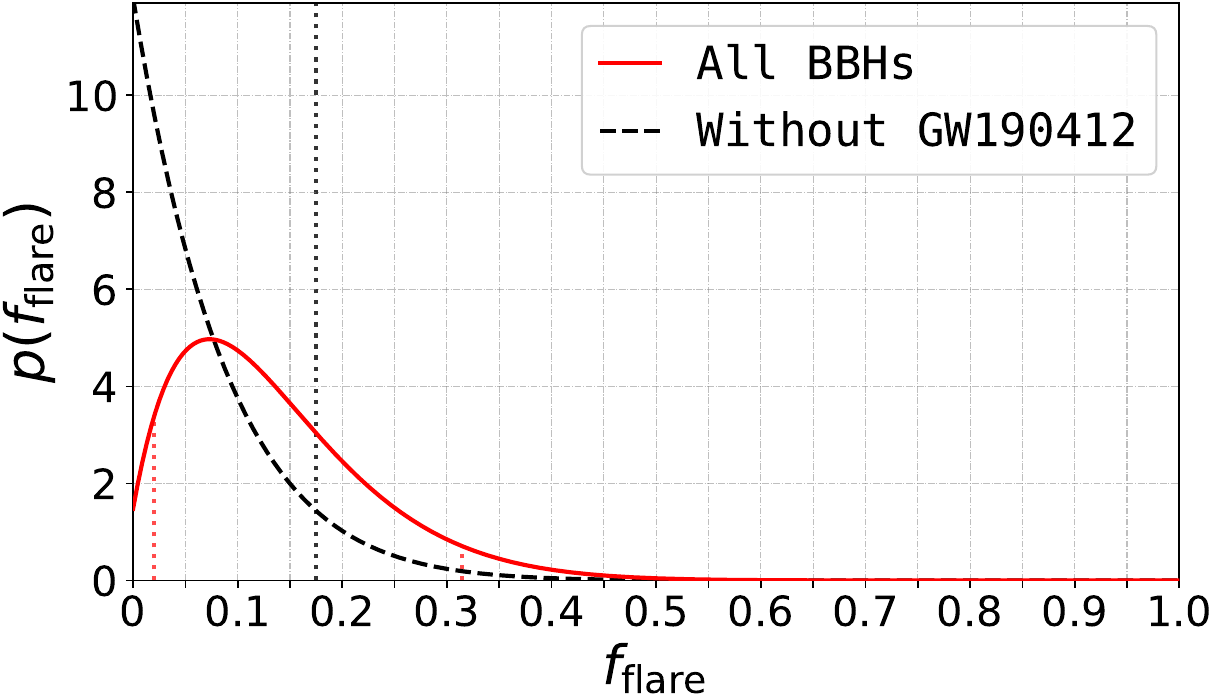}
 \caption{Probability density distributions of $f_{\rm flare}$ inferred from \citet{zhuConstrainingFractionLIGO2026}. 
 The red solid and black dashed curves represent the results derived from the all BBH events and 
 the set without GW190412, respectively. The vertical red and black dotted lines indicate 
 the error interval and the upper limit at the 90\% confidence level, correspondingly. 
 }
 \label{fig:fagnFlare_infer}
\end{figure}

A preliminary indication of a non-zero BBH--AGN-flare association was reported by \citet{zhuConstrainingFractionLIGO2026}, 
who performed a spatiotemporal correlation analysis using 80 BBH mergers from GWTC-4.0 and six years of data from ZTF DR23. 
In contrast to earlier analyses that relied on a relatively small set of unusual AGN flares 
identified by \citet{grahamLightDarkSearching2023}, this study used the much larger AGN flare catalogs 
constructed by \citet{heSystematicSearchActive2025}, including 28,504 events in the coarse flare catalog 
and 1,984 high-confidence flares in the refined catalog. 
As shown in Figure~\ref{fig:fagnFlare_infer}, they inferred a flare-associated fraction of $f_{\rm flare}=0.07^{+0.24}_{-0.05}$ at $90\%$ confidence, 
yielding a maximum-likelihood value that is non-zero. However, this indication is primarily driven by 
a single candidate flare---J143041.67+355703.8---that is spatially and temporally coincident with GW190412. 
The candidate, which is based on only two photometric measurements at the peak of its light curve, 
cannot yet be considered a confirmed electromagnetic counterpart.
When GW190412 is excluded, the inferred fraction becomes consistent with zero, 
with an upper limit of $f_{\rm flare}<0.17$ at $90\%$ confidence. 
Therefore, the result should be interpreted as a preliminary indication 
rather than a statistically established detection of a BBH--AGN-flare correlation. 

The different conclusions reached by \citet{veronesiAGNFlaresCounterparts2024}, \citet{cabreraMultimessengerConstraintsLIGO2026} 
and \citet{zhuConstrainingFractionLIGO2026} can be understood largely in terms of the flare catalogs 
and the selection of transient candidates used in the analyses. 
Both \citet{veronesiAGNFlaresCounterparts2024} and \citet{cabreraMultimessengerConstraintsLIGO2026} relied on 
the relatively restrictive catalog of 20 unusual AGN flares compiled by \citet{grahamLightDarkSearching2023}, 
which was designed to identify particularly conspicuous transient candidates and, after additional 
selection cuts, contains only a small number of events suitable for EM-counterpart searches. 
Consequently, these analyses were sensitive primarily to a small, bright, 
and highly selected subset of all AGN flaring activity. 
In contrast, \citet{zhuConstrainingFractionLIGO2026} used the six-year ZTF DR23 flare catalogs 
of \citet{heSystematicSearchActive2025}, which contain orders of magnitude more candidate flares 
and therefore provide a substantially denser sampling of the underlying AGN-flare population. 
The larger catalog increases the chance of identifying faint or otherwise less conspicuous flares that 
may be associated with BBH mergers but would not satisfy the stringent selection criteria of the earlier catalog. 
At the same time, the much larger background population also makes the treatment of chance coincidences 
and catalog completeness essential. 
The non-zero result of \citet{zhuConstrainingFractionLIGO2026} is therefore not necessarily inconsistent 
with the earlier null results: it may reflect the greater completeness of 
the newer flare catalog to a broader population of AGN variability. 
Nevertheless, because the apparent excess is dominated by a single candidate associated with GW190412, 
substantially larger and better-characterized flare samples will be required to establish 
whether this preliminary indication represents a genuine BBH--AGN connection. 

\section{Theoretical model for electromagnetic counterpart of binary black hole merger in AGN disk}
\subsection{Physical framework and general picture}
BBH mergers are widely considered to be electromagnetically dark: in isolated gas-poor environments, the inspiral and plunge eject essentially no mass, and the accretion of a tenuous ambient medium is ineffective to release a measurable amount of energy to drive observable EM signature. Thus, while BBH mergers constitute the primary gravitational wave \citep{abacGWTC50ObservationsSecond2026}, the absence of an EM counterpart renders them effectively silent in the multimessenger era. 

\begin{figure}
\centering
\includegraphics[width=0.9\linewidth]{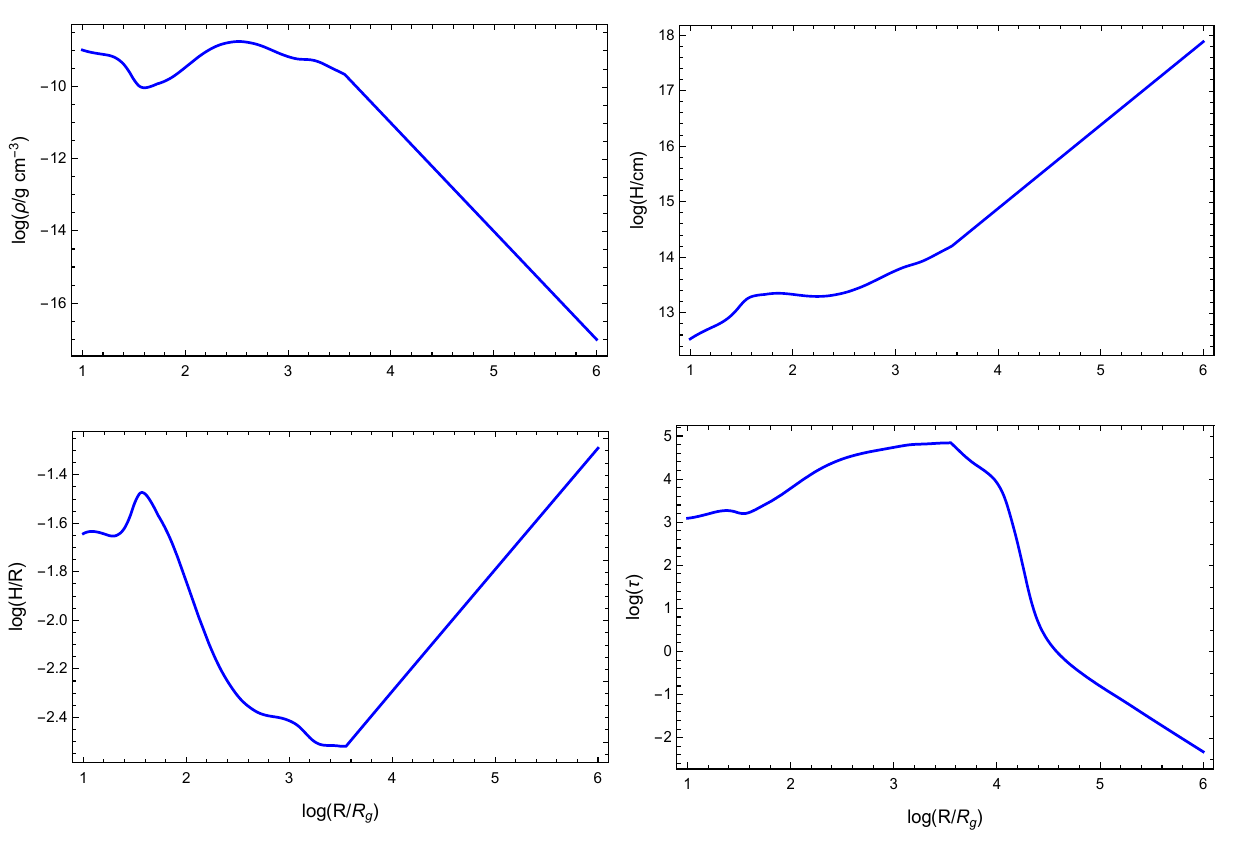}
\caption{Example radial structure of an AGN disk around a  $10^8 M_\odot$ supermassive black hole (SMBH), computed with the fiducial viscosity parameter as $\alpha=0.1$ and the SMBH accretion rate as $\dot{M}=0.1(\dot{M}_{\rm{Edd}}/\eta)$, where the radiative efficiency $\eta=0.1$. The panels show the disk gas density $\rho$, vertical thickness $H$, aspect ratio $H/R$, and optical depth $\tau$ as functions of radius. Calculations are performed using the public code \emph{pAGN} \citep{gangardtPAGNOnestopSolution2024}. }
\label{fig:agn-disk-structure}
\end{figure}

The paradigm changes fundamentally for BBH mergers embedded in the accretion disks of AGN. These disks, with high gas densities (far exceeding those of the ambient interstellar medium), large vertical thickness, and extreme optical depths \citep{sirkoSpectralEnergyDistributions2003,thompsonRadiationPressureSupported2005} (see Figure \ref{fig:agn-disk-structure} for an example AGN disk structure),
serve as a gas-rich reservoir that can be accreted by either binary component or the post-merger remnant. A stellar-mass black hole (sBH) embedded in such a disk can capture ambient gas at rates reaching \(\dot{M}_{\text{cap}}=10^{23}-10^{26}\g \s^{-1}\) 
\citep{wangAccretionmodifiedStarsAccretion2021a,tagawaCanStellarmassBlack2022,chenRoleOutflowFeedback2023}—values that far exceed the Eddington accretion rate \(\dot{M}_{\rm{Edd}}=4\pi G m m_p/(c\sigma_T)=1.4\times10^{17} (m/M_\odot)\g \s^{-1}\), where \(m\) is the mass of sBH. AGN disks have been proposed as a significant formation channel for BBH merger \citep{mckernanIntermediateMassBlack2012,yangHierarchicalBlackHole2019,tagawaFormationEvolutionCompactobject2020,mckernanMcFACTSTestingLVK2025,tagawaPropertiesBlackHole2026,vaccaroAGNdrivenBBHMergers2026}. During the late inspiral or after the merger, provided that the gravitational binding energy of the infalling gas is effectively released and the remnant's rotational energy is successfully extracted, a luminous EM counterpart can outshine the AGN background \citep{bartosRapidBrightStellarmass2017,stoneAssistedInspiralsStellar2017,mcpikeMcFACTSIVElectromagnetic2026}, thereby transforming a previously dark merger into a multimessenger event. The plausibility of these theoretical considerations is received a significant boost from the GW190521 event, for which a candidate EM counterpart—an anomalous optical flare—was reported \citep{grahamCandidateElectromagneticCounterpart2020}. This association not only lends credence to an AGN-disk origin for the merger but also suggests that BBH merger in dense disks can produce observable EM signals. Since then, a surge of theoretical and observational efforts has been devoted to exploring EM counterparts from AGN-disk-embedded black hole binaries, as we review below.

Current theoretical efforts to model EM counterparts from AGN-disk-embedded BBH mergers center around two distinct physical scenarios. The first posits that accretion from the pre-merger circum-binary disk onto the remnant BH can power the observed EM emission. The second attributes the counterpart to the recoiling BH's subsequent motion through the AGN disk, where its dynamical interaction with the ambient gas dissipates energy into observable radiation. In both scenarios, the enormous energy release ultimately derives from hyper-Eddington accretion onto the sBH—accretion that surpasses the Eddington rate by several orders of magnitude. This energy can be channeled into two forms of outflow: a wide-angle wind driven from the circum-sBH accretion disk, and a polar jet launched from the immediate vicinity of the BH. 

\subsection{Central engine and Outflows}
A hyper-Eddington accretion disk is well described by an adiabatic inflow–outflow thick-disk solution \citep{begelmanRadiativelyInefficientAccretion2012}. Inside the trapping radius \citep{begelmanCanSphericallyAccreting1979,katoBlackHoleAccretionDisks2008}, where the photon vertical diffusion timescale exceeds the radial advection timescale of the disk gas, radiation pressure efficiently drives a wind outflow, a result that has been robustly demonstrated by numerical simulations \citep{yangTwodimensionalNumericalSimulations2014,sadowskiGlobalSimulationsAxisymmetric2015,sadowskiPowerfulRadiativeJets2015,kitakiOriginsImpactOutflow2021,zhangRadiationGRMHDModels2026}. 
Ignoring the detailed wind-formation process, the wind power can be parametrized as
 \begin{equation}\label{Lw}
 L_{\text{w}}=\eta_{\text{w}} \dot{M}_{\text{cap}} c^2,
 \end{equation}
where $\eta_{\text{w}}$ is the overall efficiency of converting the captured mass into wind kinetic energy, encapsulating both the radius-dependent mass fraction of the captured gas that is ejected and the specific energy of the outflow. 

For a BBH merger, the remnant characteristically acquires dimensionless spins of order \(a_{\text{sBH}}\sim 0.7\) \citep{pretoriusEvolutionBinaryBlackHole2005}. If a large-scale magnetic flux threads the BH, its rotational energy can be efficiently extracted via the Blandford–Znajek (BZ) mechanism \citep{blandfordElectromagneticExtractionEnergy1977}, producing a collimated jet. The jet power is estimated as \citep{davisMagnetohydrodynamicsSimulationsActive2020}
\begin{equation}\label{Lj}
L_{\text{j}}=\eta_{\text{j}} \dot{M}_{\text{cap}} c^2,
\end{equation}
where $\eta_{\text{j}}$ is the jet efficiency, containing uncertainties in the BH spin, the strength of the black hole magnetic flux, and the fraction of the initial capture rate that ultimately reaches the BH. 

Although the outflow morphologies differ—a quasi-spherical wind versus a collimated jet—Equations \eqref{Lw} and \eqref{Lj} reveal a unified picture: in both cases, the outflow power scales with the initial mass capture rate, with the detailed physics subsumed into an effective efficiency factor \(\eta_{\text{W/j}}\) whose value depends on the specific model assumptions. This parametrization allows the theorists to treat the outflow—regardless of its specific launch mechanism—as the input of a central engine, effectively decoupling the energy injection from the subsequent interaction with the ambient AGN (disk) medium. By modeling this interaction—in particular the shock dissipation and radiative cooling of the outflow as it propagates through the dense environmental gas—one can compute the expected EM signatures and thus make testable predictions for the counterparts of AGN-disk-embedded BBH mergers.

\subsection{In situ and recoil-driven scenarios}
In the in situ accretion scenario, the post-merger remnant BH directly accretes gas from the pre-existing circum-binary disk, producing an outflow and releasing sufficient energy to power the EM counterpart. Pioneering this scenario, \citet{wangAccretionmodifiedStarsAccretion2021} proposed a comprehensive model for the EM counterpart of AGN-disk embedded BBH mergers, systematically predicting its multi-wavelength spectral energy distribution and light-curve evolution. In their framework, hyper-Eddington accretion onto the high-spin merger remnant both launches a BZ jet with efficiency \(\eta_{\text{j}}=10^{-5}\) and drives the so-called "Bondi explosion" wind with \(\eta_{\text{w}}=10^{-3}\) \citep{liuAccretionmodifiedStarsAccretion2024}. After breaking out of the AGN disk, both the jet and the wind expand and interact with the broad-line region of the AGN, generating thermal and non-thermal radiation simultaneously and eventually driving a giant flare spanning radio to TeV energies on timescales of months to years. Subsequent studies have improved and extended the jet-dominant picture. For an advection-dominated hyper-Eddington disk, magnetic flux can be continuously accumulated near the central sBH \citep{tagawaCanStellarmassBlack2022}, eventually forming a magnetically arrested disk (MAD) state \citep{tchekhovskoyEfficientGenerationJets2011}, which yields a substantially higher jet efficiency \(\eta_{\text{j}}=a_{\text{sBH}}^2\). Building on this, \citet{tagawaObservableSignatureMerging2023} provided a detailed description of the jet dynamical propagation through the AGN disk and the resulting cocoon structure, and predicted that the shock at the jet head during breakout from the disk surface can produce both thermal and non-thermal emission, manifesting as a peculiar transient from infrared through optical to X-ray bands. Beyond the jet breakout emission, \citet{tagawaShockCoolingBreakout2024} further proposed that the shock cooling emission of the expanding cocoon—after it escapes the optically thick AGN disk—can produce a luminous optical flare. Most recently, \citet{tagawaElectromagneticFlaresCompactobject2026} suggested that the accretion flow around the merger remnant may instead operate in an envelope mode rather than a disk mode. In this regime, the wind outflow is severely suppressed, leading to an extremely high accretion rate onto the BH and consequently a more powerful jet, with correspondingly brighter EM signatures.

The recoiling BH scenario arises from an inevitable consequence of any asymmetric BBH merger: during the final inspiral and plunge, the anisotropic emission of gravitational wave imparts a recoil kick to the merger remnant. The magnitude of this kick, \(v_{\text{k}}\), depends sensitively on the binary mass ratio and the component spins, and can scale up to \(O(10^3)\km\s^{-1}\) \citep[e.g.,][]{campanelliMaximumGravitationalRecoil2007, campanelliLargeMergerRecoils2007}. For the well-studied GW190521 event, the inferred kick is estimated to exceed \(200\km\s^{-1}\) \citep{abbottPropertiesAstrophysicalImplications2020}. As the recoiling BH traverses the AGN disk, its gravitational perturbation disturbs and captures ambient gas, potentially dissipating energy into observable channels. \citet{mckernanRampressureStrippingKicked2019} first proposed that the gas initially gravitationally bound to the kicked BH—co-moving with the remnant—undergoes ram-pressure stripping upon collision with the disk gas, producing a UV/optical flare. However, \citet{grahamCandidateElectromagneticCounterpart2020} pointed out that the bulk velocity of this bound gas is \(\sim v_{\text{k}} \ll c\), rendering the kinetic energy available for dissipation too low to power a flare that can outshine the AGN background. Instead, they assumed that the energy released in the shock tail associated with Bondi–Hoyle–Lyttleton (BHL) accretion \citep{antoniEvolutionBinariesGaseous2019} can be efficiently reprocessed into radiation, with a luminosity \(L_{\text{BHL}}=\eta \dot{M}_{\text{BHL}} c^2\), leading to a bright EM signal. 

The mass capture rate of the kicked remnant in BHL formalism is given by \citep{edgarReviewBondiHoyle2004}
\begin{equation}
\dot{M}_{\rm{cap}}= \dot{M}_{\rm{BHL}}=\frac{4 \pi G^2 m^2 \rho}{v_{\rm{k}}^{3}}=7.1\times10^{24}\g\s^{-1}\left(\frac{m}{100M_\odot}\right)^2\left(\frac{\rho}{10^{-10}\g\cm^{-3}}\right)\left(\frac{v_{\text{k}}}{10^{7.5}\cm\s^{-1}}\right)^{-3},
\end{equation}
where \(\rho\) is the density of the AGN disk. In such hyper-Eddington accretion systems, as argued above, an outflow rather than radiation serves as the primary energy carrier. Adopting an isotropic wind-like outflow, \citet{kimuraOutflowBubblesCompact2021} (with \(\eta_{\text{w}}=0.035\)) and \citet{rodriguez-ramirezOpticalUVFlares2025} (with \(\eta_{\text{w}}=0.005\)) predicted that the breakout of the outflow from the AGN disk and its subsequent expansion produce detectable soft X-ray and UV/optical flares, respectively. Additionally, \citet{maElectromagneticFlaresAssociated2025} successfully fitted the anomalous optical AGN flare associated with GW190521 \citep{grahamCandidateElectromagneticCounterpart2020} using a strong wind-outflow with \(\eta_{\text{w}}=0.07\).

\begin{figure}
\centering
\includegraphics[width=0.6\linewidth]{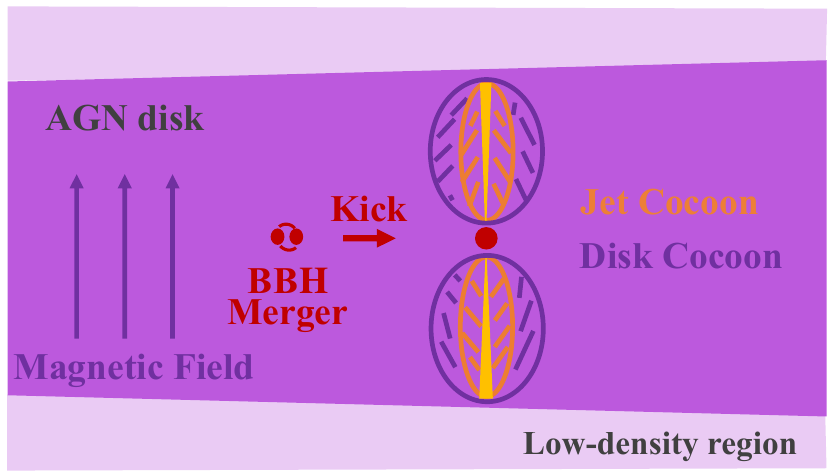}
\caption{Schematic illustration of the jet–cocoon system driven by a recoiling BBH merger remnant traversing the AGN disk. The remnant BH accretes magnetized ambient gas and launches a relativistic jet. As the jet propagates through and interacts with the disk medium, it forms a two-component cocoon structure that eventually powers the observed EM transient. }
\label{fig:ske-kick-Em}
\end{figure}

\begin{figure}
\centering
\includegraphics[width=1\linewidth]{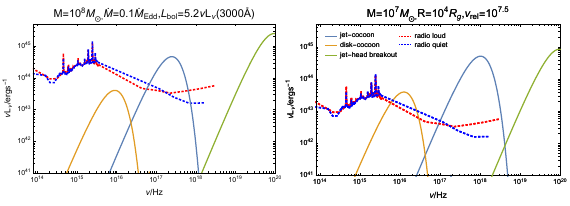}
\caption{Spectra of the emission components driven by the recoiling BBH merger remnant: the breakout jet-head  (green line), the cooling disk cocoon (golden line), and the cooling jet cocoon (navy-blue line). The left and right panels correspond to a recoiling BH of mass \(100M_\odot\) traversing the AGN disk at \(700R_g\) and \(10^4R_g\) around a \(10^8M_\odot\) and \(10^7M_\odot\) SMBH, respectively, both with a kick velocity \(v_{\text{k}}=10^{7.5}\cm\s^{-1}\). For reference, the AGN background spectra—shown as red and blue dotted lines—are constructed using the mean spectral energy distribution of radio-loud and radio-quiet quasars from \citet{shangNextGenerationAtlas2011}, with the bolometric luminosity scaling \(L_{\rm{AGN,bol}}=\zeta\lambda L_{\lambda}\) and \(\zeta(3000\mathring{A})=5.2\) \citep{runnoeUpdatingQuasarBolometric2012}. }
\label{fig:kick-SED}
\end{figure}

The recoiling BH within a realistic AGN disk environment is expected to launch a jet as a natural outcome of accretion. AGN disks are known to be magnetized \citep{kingAccretionDiscViscosity2007,salvesenAccretionDiscDynamo2016}. Drawing on numerical simulations \citep{kaazJetFormation3D2023,kimGeneralRelativisticMagnetized2025}, \citet{chenElectromagneticCounterpartsPowered2024} proposed that as the recoiling BH accretes magnetized gas from the ambient disk, magnetic flux continuously accumulates around the BH; once the system reaches a MAD state, a strong relativistic jet can be launched via the BZ mechanism (see Figure \ref{fig:ske-kick-Em}). Adopting a jet efficiency \(\eta_{\text{j}}=0.05\) inferred from the same simulations, the authors rigorously calculated the jet's propagation through the AGN disk: the jet decelerates at its head due to interaction with the dense ambient gas, while the shocked disk and jet materials are deflected laterally, forming a two-component over-pressurized cocoon that envelops the jet body \citep{matznerSupernovaHostsGammaray2003,brombergPropagationRelativisticJets2011}. In this model, three thermal radiation channels emerge: emission during jet breakout from the disk, and cooling emission from the expansion of disk cocoon and the jet cocoon after they escape the optically thick disk. Among these, the jet-cocoon cooling emission is most promising for detection, as shown in Figure \ref{fig:kick-SED}, manifesting as a soft X-ray transient lasting \(O(10^3)\s\) with a time delay \(O(10)\) days after the GW trigger. Adopting the same jet-launching mechanism but considering a jet direction nearly parallel to the disk plane, \citet{rodriguez-ramirezOpticalEmissionModel2023} found that the cooling emission from the disk cocoon can produce a detectable optical flare.

\subsection{Jet propagation and multiband emissions}
\begin{figure}
\centering
\includegraphics[width=0.99\linewidth]{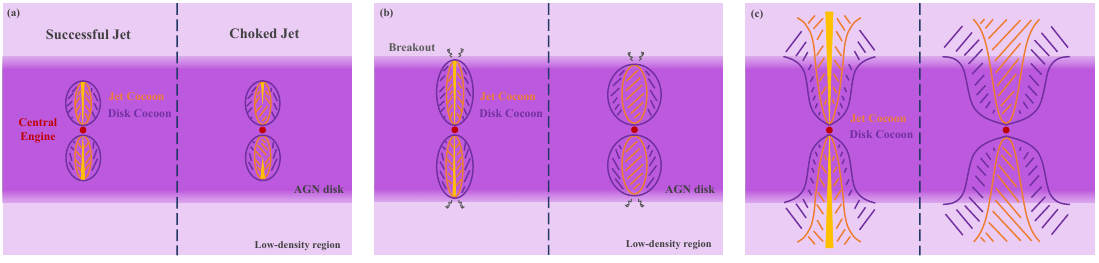}
\caption{Schematic description of jet propagation in the AGN disk. (a) Depending on the central engine duration, the jet either breaks out of the disk or becomes choked. Before engine cutoff and the jet’s tail catching its head, both cases exhibit similar structure: a jet body enveloped by a two-component cocoon. (b) In the successful case, ram pressure at the head drives rapid forward propagation, keeping the system slender. In contrast, once the jet is choked and disappears, the dissipated cocoon expands isotropically under internal pressure, resulting in a wider structure. Photons escape from the jet head or cocoon head at breakout. (c) In both cases, the two-component cocoon expands beyond the disk and continuously emits thermal radiation. }
\label{fig:ske-jet}
\end{figure}

Both the in situ accretion and recoiling BH scenarios invoke a relativistic jet as a candidate central engine powering the EM counterpart, though its power and duration are highly model-dependent. To unify previous disconnected treatments, \citet{chenObservationalPropertiesThermal2025} performed a comprehensive investigation of jet propagation in AGN disks and the associated thermal emission signatures, spanning a broad environmental parameter space (SMBH mass and disk radius). For a given jet with specific power and duration, the model determines whether the jet is choked by the AGN disk or successfully breaks out (see Figure \ref{fig:ske-jet}), and predicts the corresponding multi-wavelength EM signatures. Adopting the BHL accretion timescale \citep{kaazJetFormation3D2023},
\begin{equation}
t_{\rm{BHL}}=\frac{2G m}{v_{\rm{k}}^{3}}=8.4\times10^{5}\s\left(\frac{m}{100M_\odot}\right)\left(\frac{v_{\text{k}}}{10^{7.5}\cm\s^{-1}}\right)^{-3},
\end{equation}
as a fiducial jet duration, the authors found that remnant-launched jets can generally break out of typical AGN disks. In terms of observation, the soft X-ray flares arising from jet-cocoon cooling emission are the most prominent, with durations ranging from \(O(10^2)\s\) to \(O(10^5)\s\), provided that the jet power is not extremely low. For powerful jets (\(L_{\text{jet}}>10^{46}\erg\s^{-1}\)), detectable UV/optical flares lasting from several days to tens of days are also expected.

\begin{figure}
\centering
\includegraphics[width=0.9\linewidth]{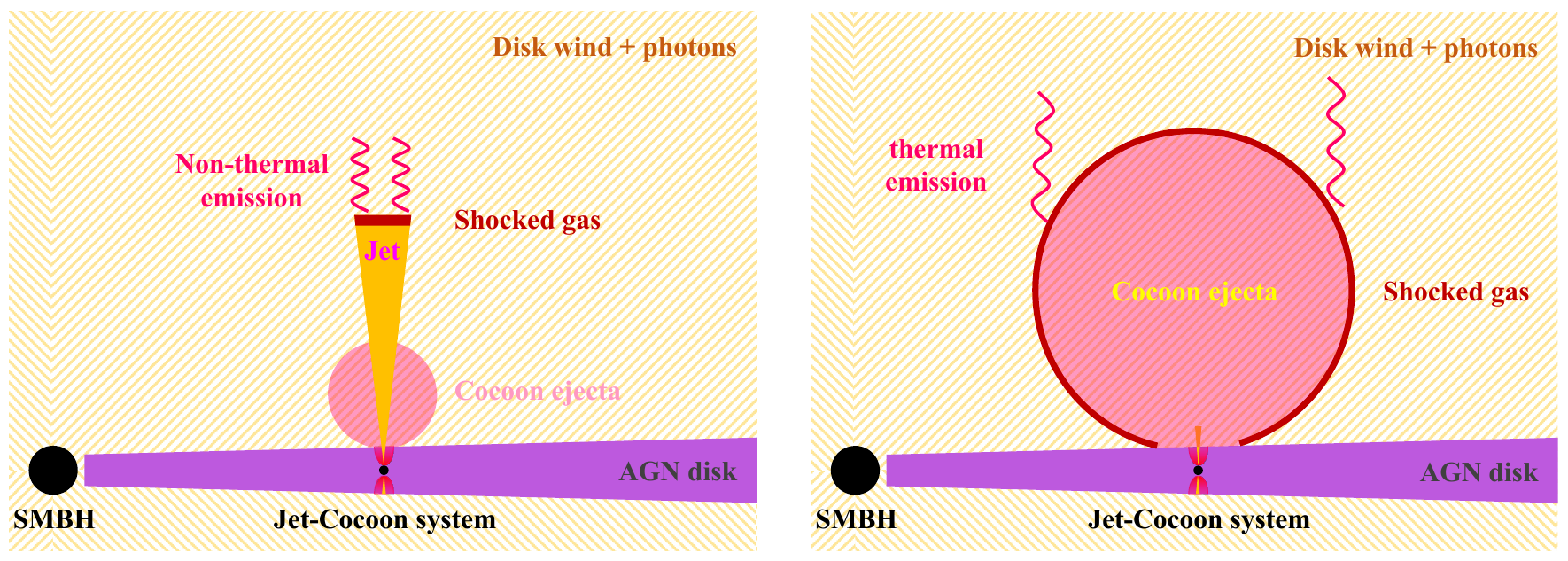}
\caption{Schematic of the two primary radiation channels from a successful breakout jet in an AGN environment, which includes the disk, wind outflows, and the ambient photon field. Strong shocks driven by jet–medium interaction accelerate particles, producing nonthermal radiation, while the expanding cocoon—pressurized by the jet's earlier propagation inside the disk—cools and emits thermal radiation.}
\label{fig:ske-BBH-EM}
\end{figure}

\begin{figure}
\centering
\includegraphics[width=1\linewidth]{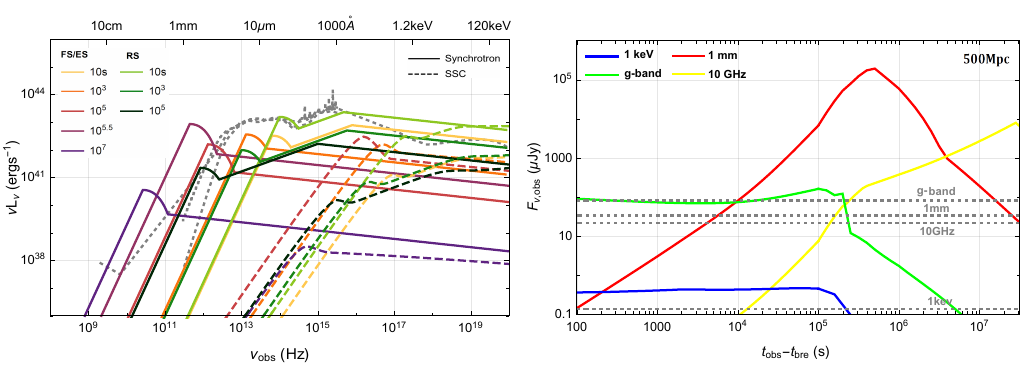}
\caption{Spectral energy distributions (forward and reverse shock) and multi-wavelength light curves of nonthermal emissions from a BBH remnant-driven jet. The results correspond to a relativistic jet with $L_{\text{j}}=10^{46}\erg\s^{-1}$ and $t_{\text{j}}=10^5\s$, consistent with the jet launched in a BBH merger system with remnant mass $m=100M_\odot$, traversing the AGN disk at \(10^3R_g\) around a \(10^7M_\odot\) SMBH with kick velocity $v_{\text{k}}=10^{7.5}\cm \s^{-1}$.}
\label{fig:jet-nonthermal}
\end{figure}

Given that jets launched by the merger remnant are generally capable of breaking out of the AGN disk, they will subsequently propagate into the disk-external environment. This region, as inferred from observations of broad emission lines and blueshifted absorption lines, is populated by dense clouds and AGN outflows \citep{netzerPhysicsEvolutionActive2013,lahaIonizedOutflowsActive2021}. Beyond the thermal emission associated with jet breakout and cocoon cooling considered in earlier works, \citet{chenObservationalPropertiesNonthermal2026} performed a comprehensive study of the long-term propagation dynamics and broadband nonthermal emission of jets in a realistic AGN medium, explicitly characterized as AGN disk winds \citep{progaDynamicsLinedrivenDisk2000,kingPowerfulOutflowsFeedback2015,giustiniGlobalViewInner2019}, as shown in Figure \ref{fig:ske-BBH-EM}. Similar to its evolution inside the disk, the jet undergoes rapid deceleration and develops a strong shock at its head. However, in the optically thin AGN-wind medium, the shock becomes collisionless rather than radiation-mediated, as is the case in the optically thick disk. Within the shock, electrons are compressed and accelerated to relativistic energy, producing nonthermal emission via synchrotron radiation, synchrotron self-Compton scattering, and external inverse-Compton scattering with AGN photons. The dense AGN environment (with gas densities reaching \(10^3-10^{10} \cm^{-3}\), far exceeding those of the typical interstellar medium) gives rise to strong synchrotron self-absorption within the shock, resulting in a prominent quasi-thermal hump at low frequencies in the emission spectrum \citep{kobayashiCharacteristicDenseEnvironment2004} (see Figure \ref{fig:jet-nonthermal}); meanwhile, the nonthermal emission can outshine the AGN background across multiple bands: soft X-ray and optical flares on the jet duration timescale, as well as low-frequency (infrared and radio) transients lasting months to years, are all expected. This multi-wavelength richness makes AGN-disk-embedded BBH mergers promising EM emitters across a broad spectral range.

\subsection{Observable signatures and identification}
BBH mergers typically occur in the outer regions of AGN disks \citep{tagawaPropertiesBlackHole2026,vaccaroAGNdrivenBBHMergers2026}. Even if such mergers induce significant perturbations to the local disk structure, these disturbances are unlikely to substantially modify the accretion flow onto the central SMBH or to trigger global AGN variability on the short post-merger timescales relevant to the GW event. Consequently, the EM counterparts of these mergers should manifest as transient flares superimposed upon the quiescent AGN continuum. A key challenge in identifying these EM counterparts, however, is the presence of intrinsic AGN variability \citep{ulrichVARIABILITYACTIVEGALACTIC1997}, which can produce flare-like features that mimic a genuine counterpart, and conversely, may cause a true merger-induced flare to be mistaken as AGN intrinsic variability. To address this, current models have proposed distinguishing criteria for GW–EM association: a temporal diagnostic—the EM signal is expected to arrive within a short time delay of hours to tens of days after the GW trigger, and one or both of the following flux-based diagnostics—the flare luminosity significantly outshines the typical AGN variability amplitude, or the EM emission exhibits distinct multi-wavelength spectral and temporal evolution. Together, these criteria provide a powerful means to securely identify merger-induced EM counterparts.

A common underpinning of current models for detectable EM counterparts is hyper-Eddington accretion onto the merger remnant, which releases enormous energy and subsequently channels it into either wind outflows or relativistic jets. Since the BHL capture rate scales as \(\dot{M}_{\rm{BHL}}\propto m^2\), more massive BBH mergers are inherently more likely to generate luminous EM signals. Indeed, the candidate EM counterparts identified so far are predominantly associated with massive BBH merger events, particularly those in the so-called upper mass gap \citep{grahamCandidateElectromagneticCounterpart2020,grahamLightDarkSearching2023,cabreraSearchingElectromagneticEmission2024,heTracingLightIdentification2025,zhangLVKS241125nMassive2026,heSearchingElectromagneticCounterpart2026}. In addition,
it is worth stressing that, to date, nearly all proposed models that yield detectable EM counterparts from AGN-disk-embedded BBH mergers are post-merger in nature. This is because the pre-merger BHs have substantially lower spins and accretion rates than the post-merger remnant, producing outflows—jet \citep{tagawaObservableSignatureMerging2023} and wind \citep{chenRoleOutflowFeedback2023} that are too weak to generate a luminous EM emission that can emerge from the AGN background.

Though the in situ accretion and recoiling BH scenarios have been developed largely independently in the literature, they are not mutually exclusive in reality. Immediately after a BBH merger, the remnant BH rapidly accretes the residual circum-binary gas, launching an early outflow and powering an initial episode of EM emission. Subsequently, the GW recoil kick displaces the remnant from its original location, redistributing its surrounding gas and potentially triggering a second hyper-Eddington accretion phase that drives another outflow and produces a later EM flare. This two-phase picture implies that the EM counterpart could exhibit a double-peaked feature—or at least a temporal evolution distinct from a single transient—which could serve as a more powerful diagnostic for discriminating merger-induced flares from AGN intrinsic variability. Nevertheless, during the late inspiral stage, hyper-Eddington accretion onto the individual BHs may drive outflow feedback that clears a low-density cavity around the BBH \citep{kimuraOutflowBubblesCompact2021,tagawaCanStellarmassBlack2022,chenRoleOutflowFeedback2023}, rarefying the circum-binary disk. If this is the case, the in situ accretion phase would be significantly suppressed or even absent, leaving only the recoil-driven flare as the observable EM counterpart. In such a scenario, the kicked remnant must travel out of the cavity and re-enter the dense AGN disk before hyper-Eddington accretion can be restarted, leading to a more prolonged time delay.

\subsection{Uncertainties and observational status}
A central assumption underlying all current models is that hyper-Eddington accretion onto the merger remnant produces powerful outflow, as supported by super-Eddington simulations. However, a critical caveat arises when considering the characteristic scales involved in AGN disk environments. The outer boundary of the BH's accretion flow in an AGN disk can be much larger—for instance, as estimated by its Hill radius \citep{chenRoleOutflowFeedback2023}:
\begin{equation}
r_{\text{Hill}}=(\frac{m}{3M})^{\frac{1}{3}} R = 1.1\times 10^{15} \cm \left(\frac{m}{100M_\odot}\right)^{\frac{1}{3}}\left(\frac{M}{10^8M_\odot}\right)^{-\frac{1}{3}}\left(\frac{R}{10^4R_g}\right),
\end{equation}
where \(M\) is the SMBH mass, \(R\) is the radial location in the disk, and \(R_g=GM/c^2\)—which far exceeds the BH's gravitational radius \(r_g\). By contrast, current numerical simulations of super-Eddington accretion are typically confined to \(O(10^4)r_g\) and initialized with inflow rates of at most 
\(O(10^4)\dot{M}_{\text{Edd}}\), far below the mass capture rates expected for BHs in AGN disks. Whether the outflow energetics inferred from such simulations can be reliably extrapolated to the much larger scales characterizing AGN disks therefore remains an open question. Recent numerical efforts have pursued global simulations of BH accretion in AGN disks, but with certain inherent limitations. Large-domain simulations fail to resolve the BH's immediate vicinity where hyper-Eddington energy release is concentrated \citep{liHydrodynamicalEvolutionBlackhole2022,kaazHydrodynamicEvolutionBinary2023,rowanBlackHoleBinary2023,whitehead3DAdiabaticSimulations2025,wangSimulationBinarysingleInteractions2025,chametlaGlobalSimulationsAccretion2026}, and therefore fail to produce outflows self-consistently. Small-scale simulations \citep{kaazJetFormation3D2023,kimGeneralRelativisticMagnetized2025}, by contrast, capture the local accretion physics but are confined to domains far smaller than the BH's gas-capture scale, still requiring significant extrapolation to AGN disk scales. Bridging these two regimes remains an open challenge. Consequently, to date, all outflow properties—whether in the form of winds or jets—remain highly uncertain. It is worth stressing that, despite several candidate EM counterparts having been reported in the literature, no unambiguous detection of an EM counterpart from a BBH merger in an AGN disk has been securely established to date. This observational gap not only hampers the rigorous testing of existing theoretical models—particularly in terms of constraining the outflow parameters—but also leaves ample room for the development of new physical scenarios and novel or more improved modeling frameworks.

Despite the substantial uncertainties in current models—and even the open question of whether a detectable EM counterpart can indeed be produced—the identification of such a counterpart remains a critical diagnostic for confirming a BBH merger in an AGN disk. First, the AGN-disk channel can produce BBH mergers with anomalous properties, including large mass (particularly in the upper mass gap) \citep{xueWhatDeterminesMaximum2025}, extreme mass ratios \citep{yangHierarchicalBlackHole2019}, high spin \citep{bartosAccretionAllYou2026}, considerable precession spin \citep{tagawaEccentricBlackHole2021}, and appreciable orbital eccentricity \citep{samsingAGNPotentialFactories2022}, all of which lie beyond the reach of the standard isolated binary evolution channel. However, dynamical merger channels in star clusters (globular clusters, young star clusters, nuclear star clusters) can also generate mergers with similar anomalous characteristics \citep{samsingFormationEccentricCompact2014,rodriguezBinaryBlackHole2016,rodriguezPostNewtonianDynamicsDense2018,hoangBlackHoleMergers2018,antoniEvolutionBinariesGaseous2019}. As a result, while GW observations of anomalous system parameters may hint at an AGN-disk origin, these signatures alone are insufficient to unambiguously confirm the formation channel. Second, identifying the host AGN would directly confirm the AGN-disk origin of a BBH merger. In practice, however, the localization accuracy of LIGO/Virgo/KAGRA for GW events is typically limited to tens to thousands of square degrees, as shown in the latest Gravitational-Wave Transient Catalog, GWTC-5.0 \citep{abacGWTC50ObservationsSecond2026}, within which the error region may contain hundreds to thousands of known AGN. This severe source confusion makes it extremely challenging to unambiguously associate a given GW event with a specific AGN. The next generation of GW detectors—the Einstein Telescope \citep{punturoEinsteinTelescopeThirdgeneration2010} and Cosmic Explorer \citep{abbottExploringSensitivityNext2017}—will achieve sub-square-degree localization \cite{zhao2018}, combined with the fact that AGN are far less numerous than galaxies, offering the prospect of direct host identification. However, with science operations not expected until the 2030s to 2040s, this capability will not be available in the near term. For the foreseeable future, spatial coincidence between GW events and AGN alone will be insufficient to securely confirm an AGN-disk origin. Therefore, detecting and securely identifying an EM counterpart carries the dual function of both localizing the host AGN and certifying the BBH merger origin in an AGN disk.

\subsection{Scientific implications and additional EM channels}
Beyond certifying the AGN-disk origin, the EM counterparts of BBH mergers in AGN disks carry broader scientific significance in the multimessenger era. The GW signal provides the luminosity distance, while the EM counterpart—via its host AGN redshift—offers an independent redshift measurement, enabling a bright siren determination of cosmological parameters, in contrast to dark sirens that rely on statistical association \citep{soares-santosFirstMeasurementHubble2019,zhaoGravitationalwaveStandardSirens2026}. With EM counterparts, events such as GW190521 and GW170817 have already been used to constrain the Hubble constant \citep{liUseBinaryBlack2025}. Looking ahead, cross-correlating GW localization regions with AGN flares among hundreds of events could improve the \(H_0\) measurement to within \(10\%\) \citep{bomStandardSirenCosmology2024}. Moreover, joint GW–EM observations can test fundamental physics \citep{barackBlackHolesGravitational2019,abbottTestsGeneralRelativity2021}. Thus, the detection of EM counterparts from AGN-disk-embedded BBH mergers carries multifaceted scientific value.

We close this section by briefly noting two speculative scenarios in which BBH mergers in AGN disks could generate EM radiation, though these remain largely unexplored at the quantitative level. The first mechanism is GW-driven viscous dissipation. GWs are known to be the most powerful astrophysical radiation \citep{keitelMostPowerfulAstrophysical2017}, with peak luminosity \(L^{\text{GW}}_{\text{peak}}\sim 10^{-3}c^5G^{-1}=10^{56}\erg\s^{-1}\). As GWs propagate through disk gas, the induced gas shear motions can dissipate a fraction of GW energy to convert into heat via viscosity \citep{hawkingPerturbationsExpandingUniverse1966}, producing a transient EM signal. The process has been studied as a potential EM counterpart for supermassive black hole binary coalescence \citep{kocsisBrighteningAccretionDisk2008}. For an AGN-disk embedded BBH merger, the GWs emitted inevitably traverse the viscous disk environment, and the resulting viscous dissipation may in principle generate EM radiation. However, despite the extreme low energy conversion efficiency, \(\dot{e}_{\text{heat}}=16\pi G \eta_s/c^2e_{\text{GW}}=3.7\times10^{-17}(\eta_s/10^{10}\g\cm^{-1}\s^{-1})e_{\text{GW}}\), where \(\dot{e}_{\text{heat}}\) is the GW dissipated rate, \(e_{\text{GW}}\) is the GW energy flux, and \(\eta_s\) is the disk shear viscosity coefficient, the short duration of the BBH late inspiral and merger is far smaller than the disk turbulent timescale, such that any heat generated is smoothed out by eddies before it can be radiated \citep{liGravitationalWaveHeating2012}. Consequently, the expected flux is likely too faint to be detected against the AGN background. The second mechanism involves the radiation deficit induced by a deep gap in the AGN disk. The gravitational interaction between an embedded BH and the large-scale disk gas exerts huge torques that repel ambient gas from the BH vicinity, opening a low-density annular gap at the BH's orbital radius \citep{kleyPlanetDiskInteractionOrbital2012,kanagawaRadialMigrationGapopening2018}. This gap results in a distinct dip or break in the multi-color blackbody spectral energy distribution of the AGN disk, as the radiative contribution from the missing annulus is effectively suppressed \citep{gultekinObservableConsequencesMergerdriven2012,mckernanIntermediatemassBlackHoles2014}. As a characteristic product of the disk-merger channel, a massive BH can readily open a deep gap in AGN disks \citep{gilbaumHowEscapeTrap2025}. Therefore, detecting such gap-dip signals could serve as evidence that BBH mergers indeed occur in AGN disks, and may act as an EM counterpart—albeit one without direct temporal association with the GW signal. Furthermore, for a massive BBH merger that has already opened a gap prior to coalescence, an extreme GW recoil kick can excite a significant orbital inclination and eccentricity of the remnant. Such orbital changes may reduce the gap depth \citep{chametlaGapFormationInclined2017,nealonWarpingProtoplanetaryDisc2018,sanchez-salcedoEstimatingDepthGaps2023}, thereby modifying the gap-induced radiation deficit and potentially producing a distinctive EM signature associated with the merger event. This offers an additional observational channel for identifying AGN-disk-embedded BBH mergers.

\section{Observational searches for electromagnetic counterparts of BBH mergers}

\subsection{Early searches following the first BBH detections}
\begin{figure}[htb!]
    \centering
    \includegraphics[width=0.8\textwidth]{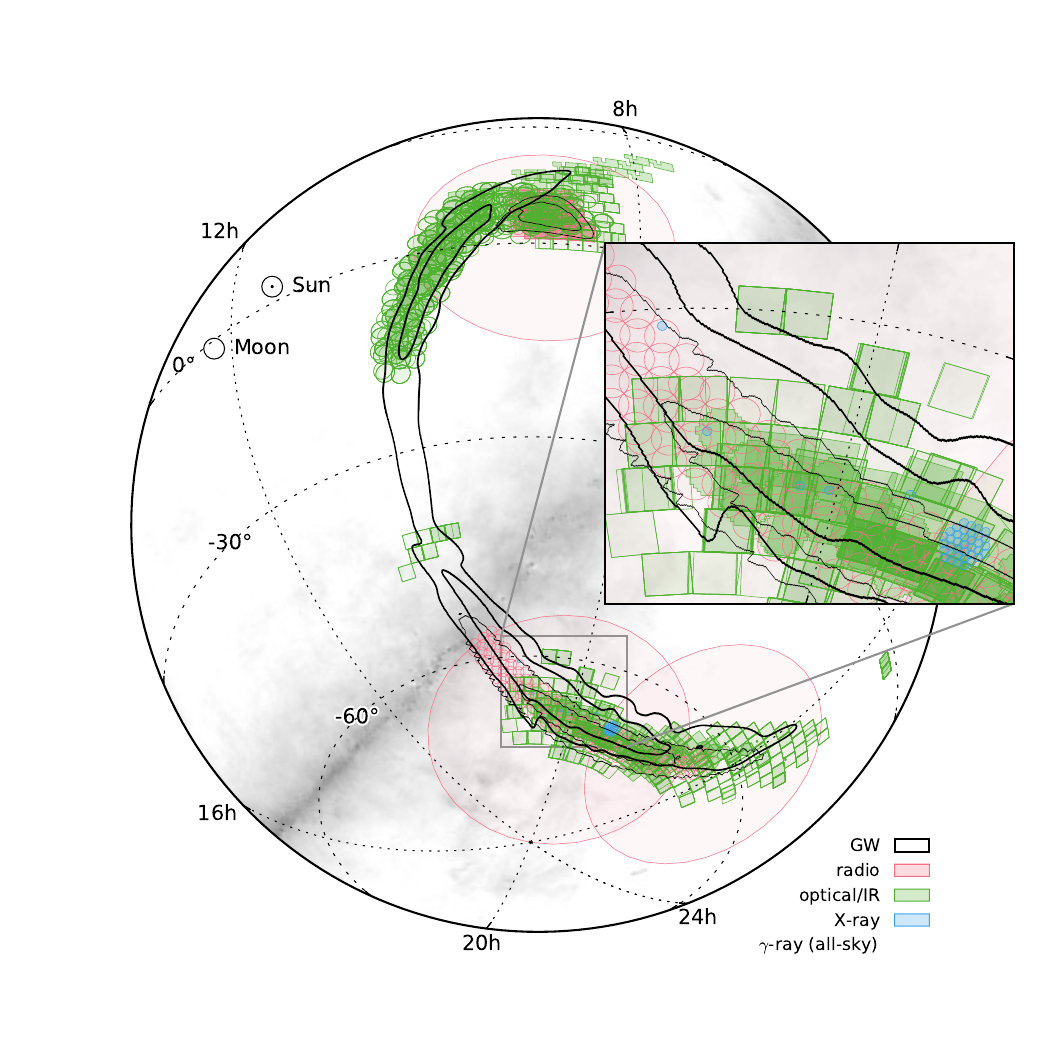}
    \caption{Footprints of observations in comparison with the 50\% and 90\% credible levels of the initially distributed GW localization maps. Adapted from Figure~2 of \citet{abbottLOCALIZATIONBROADBANDFOLLOWUP2016}.}
    \label{fig:GW150914}
\end{figure}

The search for EM emission from BBH mergers began with GW150914, the first direct detection of GWs from the merger of two stellar-mass black holes \citep{abbottObservationGravitationalWaves2016}. GW150914 triggered an extensive broadband follow-up campaign spanning gamma-ray, X-ray, optical, near-infrared, and radio wavelengths with ground- and space-based facilities \citep{abbottLOCALIZATIONBROADBANDFOLLOWUP2016,abbottSUPPLEMENTLOCALIZATIONBROADBAND2016}. As shown in Figure~\ref{fig:GW150914}, these observations provided complementary coverage of the GW localization region with facilities operating at different wavelengths.

The most notable EM candidate associated with GW150914 was reported by the \textit{Fermi} Gamma-ray Burst Monitor (GBM). A weak transient signal above 50 keV was detected approximately 0.4~s after the GW trigger and lasted for about 1~s, with a reported false-alarm probability of 0.0022 ($2.9\sigma$) \citep{connaughtonFERMIGBMOBSERVATIONS2016a}. Although the localization of the GBM signal was poorly constrained, it was broadly consistent with the GW localization region. However, no corresponding signal was detected by \textit{INTErnational Gamma-Ray Astrophysics Laboratory (INTEGRAL)} \citep{savchenkoINTEGRALUPPERLIMITS2016}. \citet{greinerFERMIGBMEVENT042016} subsequently questioned whether the GBM excess was astrophysical rather than a background fluctuation, whereas \citet{connaughtonInterpretationFermiGBMTransient2018} revisited these criticisms and maintained the originally reported significance of the candidate. The association between the GBM transient and GW150914 therefore remains unconfirmed and controversial. Nevertheless, the candidate motivated continued observational interest in the possibility of electromagnetic emission from stellar-mass BBH mergers.

Extensive searches at lower energies also found no compelling counterpart to GW150914. In the optical, wide-field surveys such as the Dark Energy Camera (DECam), the intermediate Palomar Transient Factory (iPTF) and Pan-STARRS1 identified a number of transient candidates within the GW localization region, but subsequent photometric and spectroscopic follow-up showed that they were unrelated supernovae or other background sources \citep{soares-santosDARKENERGYCAMERA2016,kasliwalIPTFSEARCHOPTICAL2016,smarttPanSTARRSPESSTOSearch2016}. Follow-up with \textit{Swift} likewise found no convincing X-ray or ultraviolet counterpart \citep{evansSwiftFollowupGravitational2016}, while radio observations with facilities including the Low Frequency Array (LOFAR), the Murchison Wide-field Array (MWA), the Australian Square Kilometer Array Pathfinder (ASKAP), and the Karl G. Jansky Very Large Array (VLA) yielded no plausible counterpart \citep{abbottSUPPLEMENTLOCALIZATIONBROADBAND2016}.

Similar multi-wavelength follow-up campaigns were carried out for subsequent BBH detections during O1 and O2. Optical surveys searched the GW localization regions for transient candidates, while gamma-ray, X-ray and radio observations provided complementary coverage over different timescales. These observations yielded predominantly non-detections and unrelated transient candidates \citep{cowperthwaiteDECAMSEARCHOPTICAL2016,smarttSEARCHOPTICALCOUNTERPART2016,evansSwiftFollowupGravitational2016a,yoshidaJGEMFollowupObservations2017,mooleyCaseStudyOnthefly2018,adrianiSearchGeVGammaRay2018,doctorSearchOpticalEmission2019,callisterFirstSearchPrompt2019,klinglerSwiftXRTFollowupGravitationalwave2019,noysenaLimitsElectromagneticCounterpart2019a,hamburgJointFermiGBMLIGO2020}. As the observing campaigns progressed, follow-up strategies benefited from increasingly mature low-latency alerting, improved GW localization, and more systematic candidate vetting. Nevertheless, no secure EM counterpart emerged from the BBH follow-up campaigns in O1 and O2.

These early follow-up campaigns were largely broad and model-agnostic. Many observing strategies had originally been developed for EM counterparts to neutron-star mergers rather than for a specific BBH emission scenario. At that time, AGNs were generally not regarded as particularly favorable environments for producing detectable BBH EM counterparts, and their intrinsic variability was often treated as a source of contamination \citep{kasliwalIPTFSEARCHOPTICAL2016,smarttSEARCHOPTICALCOUNTERPART2016,palliyaguruRADIOFOLLOWUPGRAVITATIONALWAVE2016}. In practice, pre-existing nuclear variability generally reduced the significance of a transient rather than increasing its relevance as a BBH counterpart candidate.

This perspective began to change as theoretical studies increasingly considered BBH mergers in dense, gas-rich environments, especially in the accretion disks of AGNs, where the presence of ambient gas could both facilitate the merger and provide the material required to power EM emission \citep{bartosRapidBrightStellarmass2017}. AGNs therefore became not only a source of confusing variability but also a physically motivated environment in which otherwise electromagnetically faint BBH mergers might become observable. Analyses of later O2 observations were already beginning to acknowledge this possibility \citep{gradoSearchOpticalCounterpart2020}, but systematic searches explicitly targeting AGN had not yet become an established observational strategy.

\subsection{GW190521 and its EM counterpart candidate}
A major turning point in searches for EM counterparts to BBH mergers came with GW190521. Detected by Advanced LIGO and Advanced Virgo on 2019 May 21, GW190521 was exceptional even among the rapidly growing population of BBH mergers \citep{abbottGW190521BinaryBlack2020}. Under the standard quasi-circular binary interpretation, its component masses were inferred to be $85^{+21}_{-14}\,M_{\odot}$ and $66^{+17}_{-18}\,M_{\odot}$ (90\% credible intervals). The primary component was inferred to lie in or near the mass range expected to be depleted by pair-instability supernovae \citep{fowlerNeutrinoProcessesPair1964,woosleyPulsationalPairinstabilitySupernovae2017}. Its unusual masses immediately motivated considerable interest in formation scenarios capable of producing such systems, particularly hierarchical mergers in dense stellar systems and AGN disks \citep{abbottPropertiesAstrophysicalImplications2020,tagawaMassgapMergersActive2021}.

\begin{figure}[htb!]
    \centering
    \includegraphics[width=0.7\textwidth]{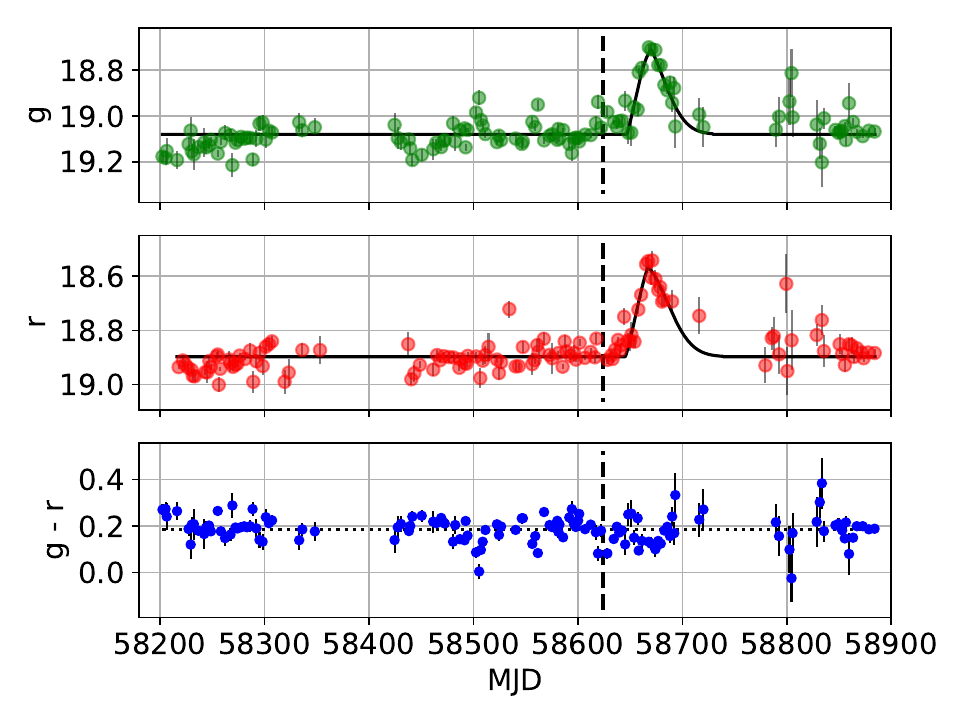}
    \caption{ZTF $g$- and $r$-band light curves of ZTF19abanrhr, the candidate EM counterpart associated with GW190521. Adapted from Figure~2 of \citet{grahamCandidateElectromagneticCounterpart2020}.}
    \label{fig:GW190521}
\end{figure}

The event subsequently attracted particular attention following the identification of a possible optical counterpart reported by \citet{grahamCandidateElectromagneticCounterpart2020}. Using observations from ZTF, they identified an optical flare, ZTF19abanrhr, associated with the previously known AGN J124942.3+344929. The AGN, at a spectroscopic redshift of $z=0.438$, lay within the GW localization region, at approximately the 78\% spatial probability contour of the skymap. ZTF covered 48\% of the 90\% localization region and identified the flare through a search for AGNs within the GW skymap that showed activity within 60 days after the GW trigger. The ZTF alert for ZTF19abanrhr was first announced approximately 34 days after GW190521. As shown in Figure~\ref{fig:GW190521}, the flare peaked $\sim 50$ days after the GW trigger, brightened by about 0.3 mag (corresponding to a luminosity enhancement of $\sim 10^{45} \mathrm{erg\, s^{-1}}$ assuming a typical quasar bolometric correction), remained elevated for $\sim 50$ days, and released a total energy of order $10^{51}$ erg. The flare was unusual relative to the historical variability of J124942.3+344929, which had varied by only a few percent over the preceding $\sim 15$ months. It also showed little significant color evolution during the flare.

\citet{grahamCandidateElectromagneticCounterpart2020} examined several possible interpretations of the flare, including intrinsic AGN variability, a supernova, microlensing, and a tidal disruption event (TDE). To quantify the likelihood of ordinary AGN variability, they modeled the stochastic variability as a damped random walk (DRW) process \citep{kellyAREVARIATIONSQUASAR2009}. Using the decade-long Catalina Real-time Transient Survey (CRTS) light curve to characterize the variability of J124942.3+344929, they generated 250,000 DRW realizations using the ZTF time sampling and found only four satisfying their flare-selection criteria, corresponding to a probability of $\sim0.002\%$ for such a flare to arise from the intrinsic variability of this particular AGN. Accounting for the look-elsewhere effect among 3255 AGNs within the 90\% three-dimensional localization region increased the estimated false-positive probability to $\sim0.5\%$. A supernova origin was disfavored by the relatively short rest-frame duration of the flare and its lack of significant color evolution. Microlensing could naturally produce achromatic variability, but its characteristic timescale is typically much longer than the several-week duration observed here, while configurations capable of reproducing the event were estimated to be rare. TDE scenarios were also considered unlikely: BH--NS disruptions were expected to be substantially more energetic and to produce a different GW signal, whereas BH--WD disruptions were generally expected to be fainter and to evolve on much longer timescales.

\citet{grahamCandidateElectromagneticCounterpart2020} further tested whether the observed properties of ZTF19abanrhr were consistent with the expected emission from a BBH merger in an AGN disk. They showed that the observed luminosity, time delay, and duration of the flare could be reproduced with physically plausible disk conditions and a recoil velocity of $\sim 200\, \rm{km\,s^{-1}}$. They also proposed two potentially testable predictions that could provide independent checks of the association. A repeat flare could occur when the kicked merger remnant reencounters the disk on a timescale of order years, while the off-center flare could produce an asymmetric broad line profile. However, the available follow-up spectrum was obtained about 200 days after the GW trigger and was therefore too late to place useful constraints on the predicted line asymmetry. 

The proposed association also motivated searches for emission at other wavelengths. \citet{podlesnyiSearchHighEnergy2020a} searched Fermi-LAT observations for $\gamma$-ray emission from the direction of J124942.3+344929, but found no significant signal in the $100\,\mathrm{MeV}$--$300\,\mathrm{GeV}$ energy range and placed upper limits on the spectral energy distribution of this source.

Despite these suggestive features, the association between GW190521 and ZTF19abanrhr remained uncertain. \citet{ashtonCurrentObservationsAre2021} reassessed the significance of the association using the three-dimensional localization overlap between the GW event and the AGN. They found that the odds in favor of a common origin over a chance coincidence ranged from only $\sim1$ to $12$, depending on the waveform model adopted for the GW analysis. They therefore argued that the available observations were insufficient to confidently associate ZTF19abanrhr with GW190521 and cautioned against drawing astrophysical conclusions that relied on the association. A complementary concern was raised by \citet{palmeseLIGOVirgoBlack2021}, who emphasized the large localization volume of GW190521 and the correspondingly high probability of chance coincidence with unrelated AGN variability. They estimated that its 90\% localization volume could contain approximately 7400 unobscured AGNs brighter than $g=20.5$ AB mag, leading to a $\gtrsim70\%$ probability of finding an AGN flare consistent with the GW event by chance. They therefore argued that a single spatially and temporally coincident flare was insufficient to establish a confident association and emphasized the need for population-level analyses of multiple BBH events.

Subsequent reanalyses, however, yielded somewhat greater support for the proposed association. \citet{bustilloGW190521BlackholeMerger2021} reanalyzed the GW signal using the \texttt{NRSur7dq4} waveform model under a different mass-ratio prior and obtained an odds ratio of $72:1$ in favor of a common origin over a random coincidence. Later, \citet{mortonGW190521BinaryBlack2023} revisited the association using the updated GWTC-2.1 \citep{abbottGWTC21DeepExtended2024} data release and a Bayesian model-selection framework that incorporated not only the sky position and luminosity distance but also the inferred primary mass and the expected mass distribution of BBHs formed in AGN disks. Adopting the astrophysical prior odds of $1/13$ used by \citet{ashtonCurrentObservationsAre2021}, they found an odds ratio of approximately $400:1$. Overall, the inferred significance remained sensitive to the GW analysis and underlying astrophysical assumptions, leaving the association suggestive rather than conclusive.  

An independent uncertainty concerns the physical origin of the optical flare itself. \citet{depaolisQuasarMicrolensingEvent2020} showed that its ZTF light curve could be well described by a standard Paczynski (single lens) microlensing profile with an Einstein crossing time of $t_{\rm E}=17.6\pm1.2$ days, and argued that a lens with a mass of $\sim0.1\,M_{\odot}$ could reproduce the observed event. In this interpretation, the nearly achromatic evolution of ZTF19abanrhr arises naturally from gravitational lensing rather than from a flare associated with the BBH merger. More recently, \citet{cazzollaRevisitingQuasarMicrolensing2026} revisited the ZTF light curve using several microlensing models and again found that the optical bump could be reproduced by microlensing, with a characteristic lens mass of $\sim0.1\,M_{\odot}$. These studies therefore demonstrate that a chance microlensing event remains a viable alternative explanation for ZTF19abanrhr, independent of the statistical consistency between the AGN position and the GW localization.

Taken together, GW190521 and ZTF19abanrhr remain one of the most prominent candidate associations between a BBH merger and an EM transient. The temporal, spatial, and phenomenological consistency of the flare with an AGN disk merger scenario makes the association particularly intriguing, but neither the statistical evidence nor the physical interpretation of the optical flare is yet definitive. Nevertheless, the candidate marked an important shift in BBH EM counterpart searches. Although intrinsic AGN variability remains a major source of background contamination, AGN flares with properties consistent with BBH merger models came to be considered physically motivated counterpart candidates rather than being rejected solely because of their association with an AGN. This development provided strong motivation for subsequent systematic searches for flares associated with BBH mergers in AGNs. 

\subsection{AGN-targeted searches and follow-up}

The proposed association between GW190521 and ZTF19abanrhr motivated an important change in the optical search strategy for BBH EM counterparts. Instead of being treated mainly as a contaminating source of variability, AGNs were increasingly recognized as physically motivated environments in which BBH mergers might produce detectable EM emission. This perspective encouraged the combination of GW localizations, AGN catalogs and wide-field time-domain surveys to search systematically for flares from AGNs within GW localization volumes. As a result, the search focus moved from identifying generic transients following a GW event to directly examining the variability of AGNs as potential counterpart hosts.

The basic strategy of these searches is to crossmatch GW skymaps with catalogs of known AGNs and then examine their optical light curves for anomalous flaring activity following the GW trigger. The key requirement is not simply that an AGN brightens after a merger, but that the observed flare represents a significant departure from its normal stochastic variability. Long-term time-domain data are therefore essential for characterizing the underlying AGN variability and identifying significant post-merger flares. Candidate flares must then be distinguished from other transient phenomena, such as supernovae, tidal disruption events, and microlensing, before being considered plausible BBH EM counterpart candidates.

\citet{grahamLightDarkSearching2023} carried out the first systematic search for BBH EM counterparts targeting AGNs across the O3 GW sample. Using ZTF observations, they searched for unusual flaring activity in AGNs potentially associated with 83 BBH and lower-mass-gap merger alerts from O3. The search targeted AGNs drawn primarily from the Million Quasars Catalog \citep{fleschMillionQuasarsMilliquas2023} that were consistent with the three-dimensional localization of the GW events. Their light curves were then examined for flares occurring within 200 days after the merger. To distinguish these flares from intrinsic AGN variability, they modeled the stochastic variability of each AGN as a damped random walk and applied a Gaussian-process change-point analysis to identify significant departures from the underlying variability. Other possible contaminants, including SNe, TDEs, and microlensing events, were further excluded based on their light curve morphology, energetics, color evolution, and other transient properties.

After these selections, 20 unusual AGN flares remained in the full ZTF data set. Seven of them were spatially and temporally consistent with one or more of nine O3 GW events. \citet{grahamLightDarkSearching2023} reported a chance coincidence probability of $p=1.9\times 10^{-3}$ for the observed associations. This result provided the first indication that AGN flares associated with GW events might occur more frequently than expected from random coincidence, although the poorly characterized population of unusual AGN flares remained an important source of false positives.

The nature of these candidate flares was subsequently re-examined by \citet{heTracingLightIdentification2025} using approximately three additional years of ZTF observations, allowing their variability to be evaluated against a much longer history of AGN activity. Only three of the original seven flares remained identifiable as distinct flare-like events in the extended light curves, while the others became less unusual in the context of their long-term variability. A joint Bayesian analysis of the optical and GW data found strong associations between two of these flares and their corresponding GW events. No similar secondary flares were identified in the host AGNs of these two candidates through 2024 October 31. This long-term reassessment highlighted the importance of extended time-domain monitoring for evaluating AGN flare candidates.

A further development was the construction of large, systematic AGN flare catalogs independent of individual GW triggers. Using six years of ZTF DR23 observations, \citet{heSystematicSearchActive2025} analyzed the light curves of a large AGN sample and constructed two catalogs of AGN flares. The AGN Flare Coarse Catalog (AGNFCC) contains 28,504 AGN flares identified through Bayesian blocks and Gaussian processes, while the more restrictive AGN Flare Refined Catalog (AGNFRC) contains 1,984 high-confidence flares. Constructing the flare sample independently of any specific GW event allows the occurrence rate, temporal properties, and contaminating populations of AGN flares to be studied before a GW association is considered. These catalogs can then be directly crossmatched with GW localization regions to identify potential BBH counterpart candidates.

\begin{figure}[htb!]
    \centering
    \includegraphics[width=0.7\textwidth]{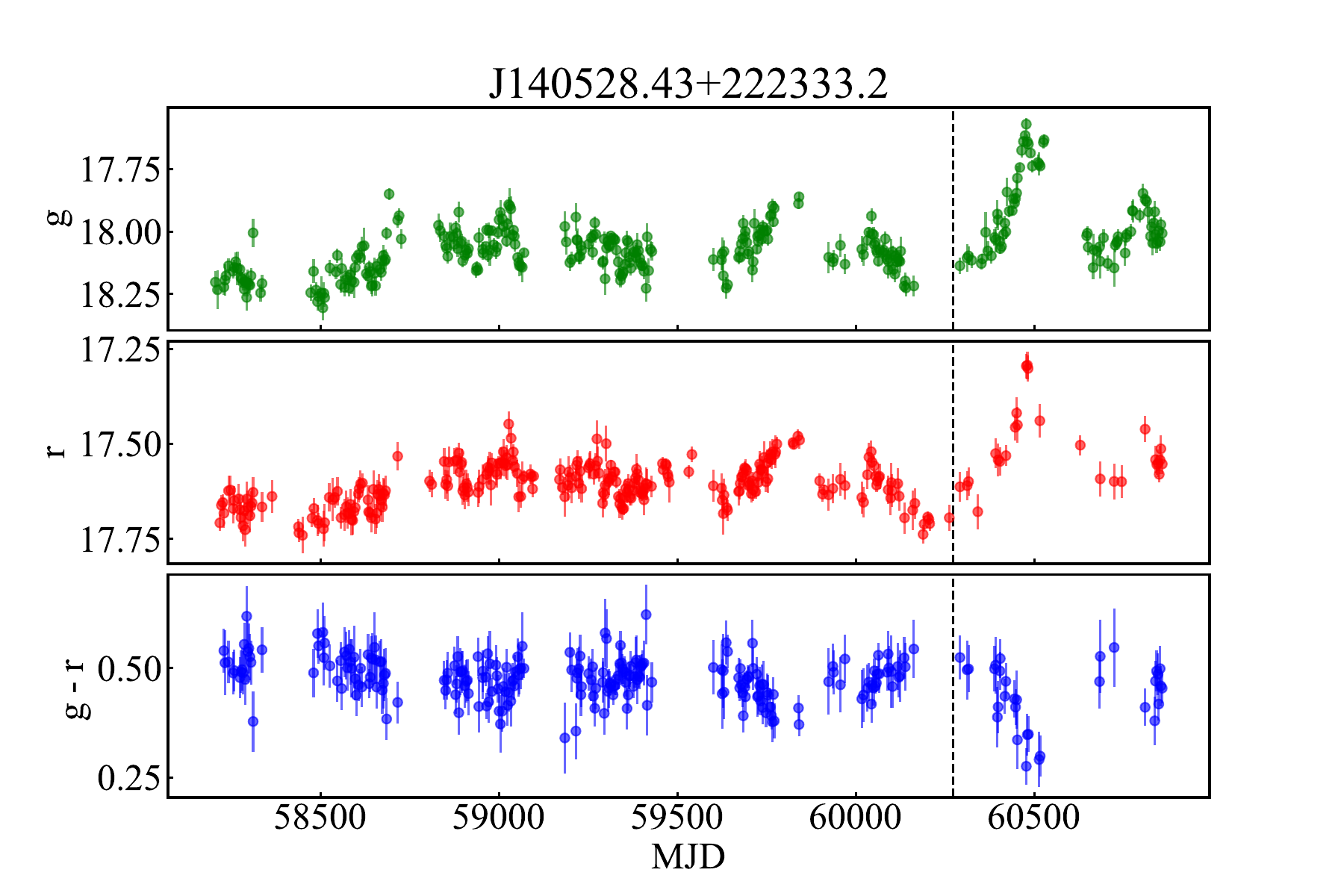}
    \caption{Light curve of one representative optical flare candidate identified in the search for an EM counterpart to GW231123. Adapted from Figure~3 of \citet{heSearchingElectromagneticCounterpart2026}.}
    \label{fig:GW231123_EM}
\end{figure}

This catalog-based strategy was subsequently applied to GW231123, the most massive BBH merger detected to date \citep{abacGW231123BinaryBlack2025a} and therefore a particularly interesting target for EM counterpart searches. \citet{heSearchingElectromagneticCounterpart2026} crossmatched its localization with the AGNFCC and identified six plausible optical flare candidates. These flares occurred after the merger, were spatially consistent with the GW localization, and represented significant departures from the stochastic variability of their host AGNs. Figure~\ref{fig:GW231123_EM}
shows the light curve of one representative candidate. None of these candidates has yet been confirmed as an EM counterpart to GW231123, but the search illustrates the utility of this strategy.

A similar archival search was performed by \citet{heExploringHierarchicalMerger2026} for GW241011 and GW241110, two asymmetric BBH mergers with properties suggestive of hierarchical formation \citep{abacGW241011GW241110Exploring2025}. Motivated by the possible contribution of AGN disks to their formation, they crossmatched the GW localizations with spectroscopically confirmed AGNs from several catalogs and examined their ZTF and ATLAS forced-photometry light curves. No convincing flare was found for GW241011, while one AGN spatially consistent with GW241110 showed a possible brightening approximately 100 days after the merger. However, the brightening was not significant compared with the long-term variability of the source and was therefore regarded only as a tentative candidate. Although no compelling counterpart was identified, the study provides another example of applying archival AGN light curves to search for EM counterparts to BBH mergers.

AGN-oriented counterpart searches have also been implemented through automated alert-broker systems. \citet{bommireddyBrokerintegratedAlgorithmElectromagnetic2026} developed an automated search for BBH optical counterparts during O4a and O4b using public ZTF alerts processed through the Automatic Learning for Rapid Classification of Events (ALeRCE) broker \citep{sanchez-saezAlertClassificationALeRCE2021}. The pipeline first selected ZTF alerts within the GW localization regions that were spatially consistent with known AGNs, and then filtered these AGN-associated transients using temporal cuts, machine-learning classifications, and host-galaxy information. The search yielded one candidate in O4a and four in O4b. This work illustrates how broker-based searches can systematically identify and assess AGN-associated transients within BBH counterpart searches.

In addition to archival searches, dedicated follow-up observations of individual GW events can also identify AGN-associated counterpart candidates. \citet{cabreraSearchingElectromagneticEmission2024} carried out long-term follow-up of the BBH merger S230922g (GW230922\_020344 in GWTC-4.0) to search for an EM counterpart, primarily using wide-field imaging from the DECam and supplemented by additional photometric and spectroscopic observations. The search covered approximately 70\% of the GW sky localization probability and discovered six potential counterpart candidates. Among them, AT~2023aagj, a transient associated with an AGN host, was considered particularly interesting because its broad line profiles exhibited temporally varying asymmetric components. Such spectral features may be consistent with an off-center transient in the AGN, providing an additional diagnostic beyond photometric flaring alone. This work demonstrated the value of spectroscopic monitoring for further characterizing AGN-associated candidates identified through wide-field GW follow-up.

A more explicitly AGN-targeted follow-up strategy was adopted by \citet{darcLongtermOpticalFollow2025} for S231206cc (GW231206\_233901 in GWTC-4.0). Using the T80-South telescope as part of the S-PLUS Transient Extension Program, they prioritized fields within the GW localization containing known AGNs and monitored these regions over an extended period after the merger. No candidate satisfied the criteria for a viable optical counterpart. Nevertheless, the nondetection was used to constrain several models for optical emission from BBH mergers in AGN disks, demonstrating that dedicated AGN-targeted follow-up can provide useful physical constraints even in the absence of a detected counterpart. 

A related follow-up study was carried out for S240413p (GW240413\_022019 in GWTC-4.0) by \citet{darcSearchingBinaryBlack2026}. Wide-field optical observations yielded a large sample of transient candidates, which were subsequently crossmatched with known AGNs, leaving two AGN-associated sources for further investigation. Long-term ZTF light curves revealed no convincing flare associated with the GW event, with the observed variability remaining consistent with intrinsic AGN activity. Spectroscopic observations were used to characterize the AGN hosts, and the inferred host properties were incorporated into an AGN disk emission model to estimate the expected delays of possible merger-induced flares. This study illustrates how long-term photometric monitoring and host characterization can be combined with physical models to evaluate AGN-associated BBH counterpart candidates.

Taken together, these studies show that searches for BBH EM counterparts in AGNs have progressed from isolated candidate associations to a broader observational effort across multiple GW events. The accumulated results provide a basis for assessing what current observations have established and what remains necessary for robust counterpart identification.

\subsection{Observational lessons and future prospects}

\subsubsection{Lessons from current observations}

More than a decade after the first direct detection of a BBH merger, no EM counterpart has yet been conclusively identified. Nevertheless, several plausible candidates have been reported. The most notable remains the proposed association between GW190521 and the optical flare ZTF19abanrhr, which has provided an influential example of what a BBH counterpart in an AGN disk might look like. Subsequent searches have identified additional AGN flares that are spatially and temporally consistent with individual GW events, extending the search beyond this single candidate association. Although none has yet provided decisive evidence for a physical connection with a BBH merger, these searches have established AGN flares as a plausible and observationally motivated class of counterpart candidates.

An important lesson from these searches is that spatial and temporal coincidence alone is generally insufficient to establish a physical association between an AGN flare and a BBH merger. Large GW localization volumes can contain many AGNs, some of which may exhibit unrelated flaring activity within the relevant time window. The significance of a candidate therefore depends not only on its consistency with the GW localization, but also on the background rate of comparable AGN flares and the number of potential hosts being searched. This issue was highlighted by reassessments of GW190521--ZTF19abanrhr, for which the inferred association significance was found to depend sensitively on the adopted GW analysis and astrophysical assumptions. More generally, long-term monitoring of previously identified candidates has shown that the apparent significance of an individual flare can change when it is evaluated against a longer history of AGN variability. Robust counterpart identification therefore requires both a well-characterized flare background and statistical analyses that account for chance coincidences across multiple AGNs and GW events.

A major methodological shift has been the changing treatment of AGNs in BBH counterpart searches. In early follow-up campaigns, variability from known AGNs was often regarded as contamination, since intrinsic AGN activity could naturally account for transients detected within GW localization regions. With the development of AGN disk scenarios for BBH mergers, however, AGNs came to be recognized as physically motivated environments for BBH mergers. As a result, searches increasingly began to retain and specifically examine transient activity associated with AGNs rather than rejecting it as background variability.

Among the different wavelength bands, optical time-domain surveys have so far played the dominant role in searching for BBH counterparts in AGN disks. Wide-field optical surveys provide both the sky coverage needed for large GW localization regions and the long temporal baselines required to characterize variability in AGNs, making them particularly suitable for searches targeting pre-existing AGNs. High-energy and radio observations, in contrast, have so far provided mainly non-detections and upper limits, which constrain possible emission scenarios but have not yet offered independent confirmation of optical candidates.

\subsubsection{Current observational challenges}

Despite these advances, establishing a physical association between an AGN flare and a BBH merger remains difficult. A central challenge is the intrinsic variability of AGNs themselves. AGNs exhibit stochastic variability over a broad range of timescales and wavelengths \citep{ulrichVARIABILITYACTIVEGALACTIC1997,kellyAREVARIATIONSQUASAR2009}. Some of this variability can resemble the brightening expected from a BBH merger in an AGN disk. Therefore, a flare occurring after a GW event may be unusual without necessarily being related to the merger. Distinguishing merger-induced emission from the normal variability of the host AGN requires a sufficiently long light curve to characterize its typical behavior. Stochastic AGN variability can be characterized with models such as the DRW \citep{macleodMODELINGTIMEVARIABILITY2010} or more general GP methods \citep{mclaughlinUsingGaussianProcesses2024}, providing a statistical baseline for assessing whether a post-merger flare is genuinely anomalous.

In addition to intrinsic AGN variability, other nuclear transients can complicate the interpretation of candidate counterparts. Supernovae can occur near the galactic centers or within AGN accretion disks \citep{grishinSupernovaExplosionsActive2021} and produce optical transients that may overlap with the expected luminosities and timescales of BBH merger flares. TDEs represent another important source of nuclear flaring and can produce luminous optical emission lasting from weeks to months \citep{gezariTidalDisruptionEvents2021}. Gravitational microlensing can also produce substantial changes in the observed brightness of an AGN without requiring any intrinsic change in its emission \citep{lawrenceSlowblueNuclearHypervariables2016}. In many cases, the overall light curve morphology alone is insufficient to distinguish among these possibilities. Candidate assessment therefore requires additional information, including color evolution and spectral properties.

Large GW localization volumes further compound this statistical challenge by containing many potential AGN hosts. The resulting chance-coincidence background makes it difficult to assign high confidence to any individual spatially and temporally consistent flare.

Counterpart searches are also limited by the incompleteness of AGN catalogs compiled from different surveys. This limitation is particularly severe near the Galactic plane, where stellar contamination and dust extinction make AGN identification substantially more difficult. As a result, some potential host AGNs within a GW localization region may simply be absent from the catalogs used for searches.

A more fundamental limitation is that detectable EM emission is expected from only a subset of BBH mergers, even under ideal observing conditions. The AGN disk channel itself contributes only a fraction of BBH population \citep{zhuEvidenceFractionLIGO2025}, and among these mergers, detectable optical emission is expected primarily from mergers in unobscured Type I AGNs and from remnants kicked toward the observer, where the emission is less affected by obscuration from the accretion disk.

Even for this favorable subset of mergers, however, the expected EM signal remains highly uncertain. Its luminosity, duration, and time delay relative to the GW event depend sensitively on the poorly constrained properties of the AGN disk and the local merger environment. Different models therefore predict a wide range of possible observational signatures, making it difficult to define a well-motivated time window or sensitivity threshold for counterpart searches.

In practice, these uncertainties make wide-field time-domain surveys the most practical way to search for such counterparts over a broad range of timescales. However, their limited depth and cadence, seasonal gaps, and incomplete sky coverage can still cause genuine flares to be missed or poorly sampled. Consequently, the success of a counterpart search depends not only on whether detectable emission is produced, but also on whether it occurs within the temporal, spatial, and sensitivity coverage of the available observations.

\subsubsection{Toward robust counterpart identification and future searches}

Given the limitations discussed above, a convincing association will likely require evidence beyond a single flare that is spatially and temporally consistent with a GW event. More distinctive observational signatures are needed to distinguish the EM counterpart emission from intrinsic AGN variability and other nuclear transients.

Repeated flares may provide a useful diagnostic in scenarios where the merger remnant undergoes multiple interactions with the AGN disk. After the merger, a recoiling remnant may encounter dense disk gas or cross the disk multiple times, potentially producing additional episodes of enhanced emission on timescales of order years \citep{grahamCandidateElectromagneticCounterpart2020}. This possibility has motivated long-term monitoring of existing candidates. For example, \citet{heTracingLightIdentification2025} found no secondary flare in the host AGNs of the O3 candidates reported by \citet{grahamLightDarkSearching2023} up to 2024 October 31. Such non-detections do not necessarily rule out an AGN disk origin, because the timing and brightness of repeated emission depend strongly on the disk structure, merger location, and recoil velocity. More importantly, the detection of a repeated flare would provide stronger evidence for an association if its delay and other observable properties were consistent with a specific post-merger process.

Spectroscopy may provide an additional diagnostic for establishing a physical association between an AGN flare and a BBH merger. In the AGN disk scenario, a recoiling merger remnant can produce an off-center flare within the disk, which may illuminate the pre-existing broad-line region asymmetrically and thereby generate asymmetric broad emission-line profiles \citep{mckernanRampressureStrippingKicked2019}. A possible example was reported for AT~2023aagj following S230922g, whose spectrum obtained near the flare peak showed asymmetries in broad emission lines, while the H$\alpha$ asymmetry was no longer apparent in a later spectrum \citep{cabreraSearchingElectromagneticEmission2024}. Although such features are not unique to BBH mergers, spectroscopy provides information on the structure and geometry of the broad line region that is unavailable from photometric variability alone, making it an important tool for assessing candidate counterparts.

Finally, multi-wavelength observations can probe different physical components of the merger environment and provide complementary constraints on the origin of a candidate. A temporally consistent counterpart detected across optical, X-ray, gamma-ray, or radio wavelengths would therefore provide a more complete picture of the underlying emission process and strengthen the case for an association with the BBH merger. Searches at high-energy and radio wavelengths have already been performed for several BBH events \citep{bhaktaJAGWARProwlsLIGO2021,ramanSwiftBATGUANOFollowup2025}. Although no coherent multi-wavelength counterpart has yet been established, coordinated observations across different bands could provide a much stronger basis for future BBH counterpart identification.

Taken together, these considerations suggest that future counterpart searches will benefit from a more coordinated observational strategy. Wide-field time-domain facilities, such as the ZTF \citep{bellmZwickyTransientFacility2019}, WFST \citep{wangScience25meterWide2023,2023ApJ...947...59L}, and the Vera C. Rubin Observatory \citep{ivezicLSSTScienceDrivers2019}, can enable the discovery of candidate flares and provide long-term photometric baselines. Promising candidates should be followed rapidly with spectroscopy and, where possible, complementary observations at other wavelengths with Einstein Probe (EP) \citep{ep}, Space-based multi-band astronomical Variable Objects Monitor (SVOM) \citep{svom}, enhanced X-ray Timing and Polarimetry mission (eXTP) \citep{eXTP}, Five-hundred-meter Aperture Spherical radio Telescope (FAST), Square Kilometre Array (SKA) and other facilities. Subsequent monitoring is also important for identifying repeated activity. Integrating these observations into a coordinated framework will be essential for moving from the identification of plausible AGN flares toward more robust associations with BBH mergers.

\subsubsection{Scientific implications of future detections}

\begin{figure}[htb!]
    \centering
    \includegraphics[width=0.49\textwidth]{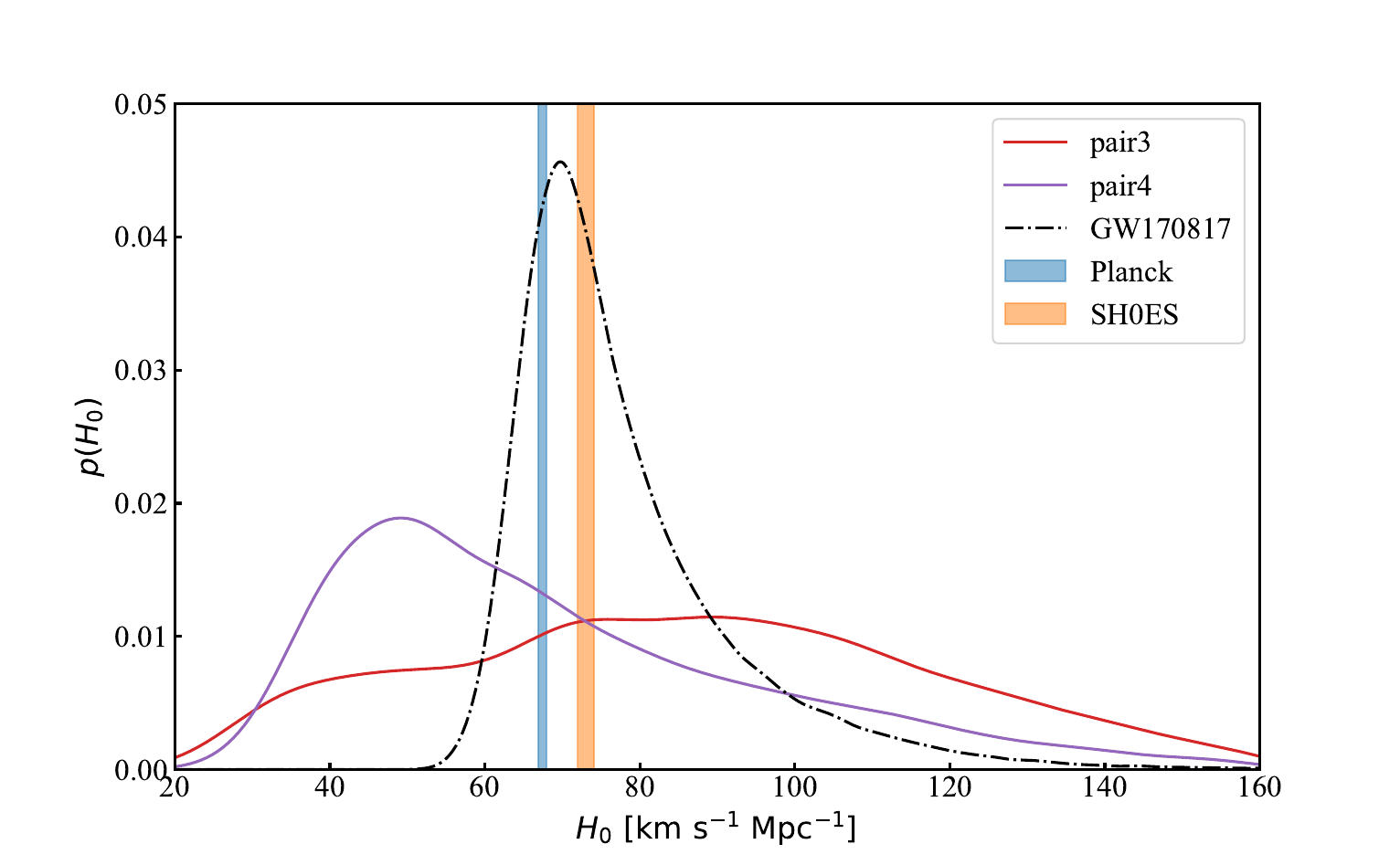}
    \includegraphics[width=0.49\textwidth]{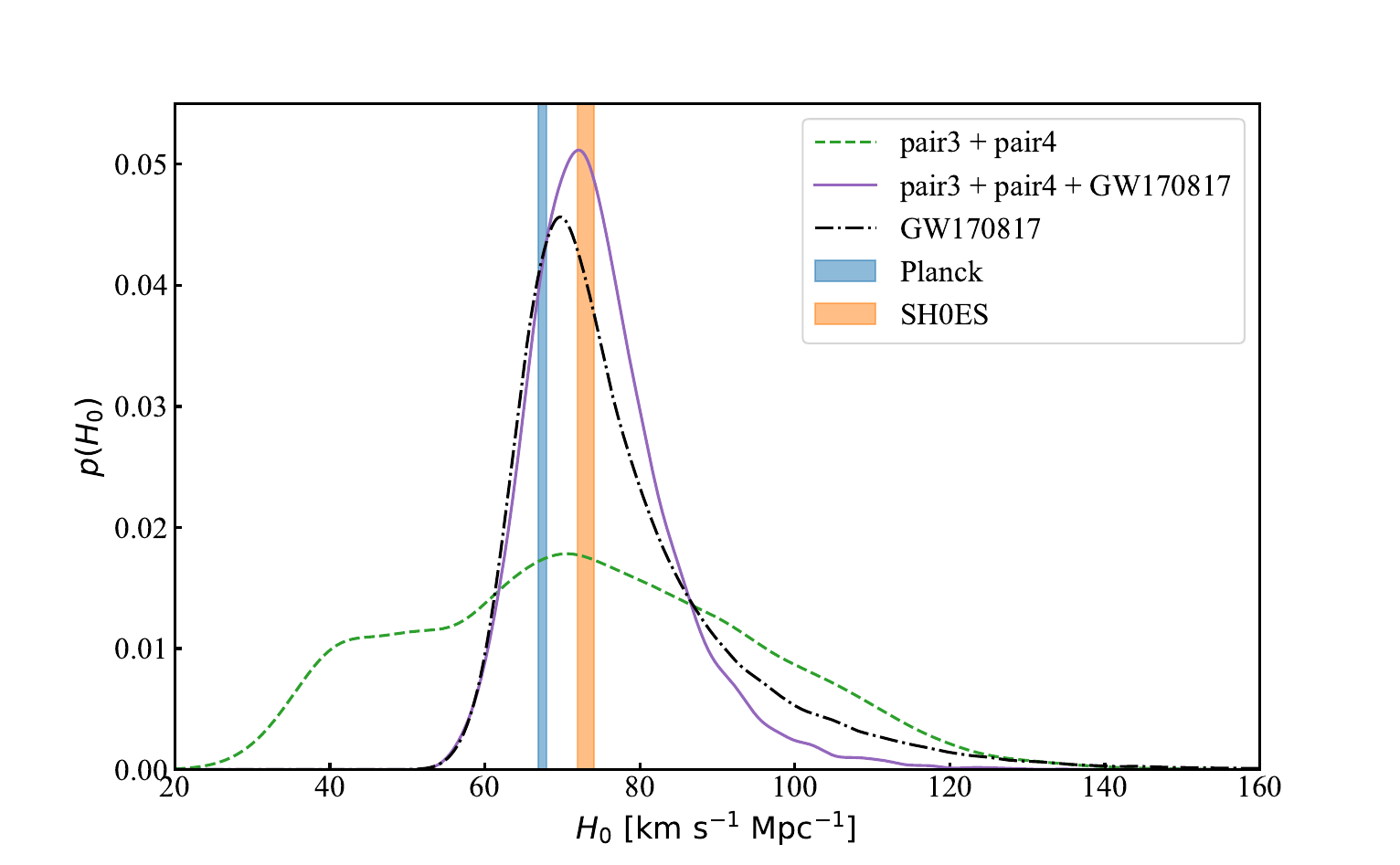}
    \caption{Constraints on the Hubble constant from the candidate associations of GW190521 with ZTF19abanrhr (pair 3) and GW190803 with J120437.98+500024.0 (pair 4). The left panel shows the individual constraints, while the right panel shows their combined constraint. The constraint from GW170817 and the 1$\sigma$ intervals from Planck and SH0ES are shown for comparison. Adapted from Figure~5 of \citet{heTracingLightIdentification2025}.}
    \label{fig:H0}
\end{figure}

A confirmed EM counterpart would enable the identification of the host AGN and provide a direct measurement of its redshift. Combined with the luminosity distance inferred from the GW signal, this would allow BBH mergers to be used as standard sirens for cosmological measurements \citep{abbottGravitationalwaveStandardSiren2017}. This possibility has already been explored for the proposed association between GW190521 and ZTF19abanrhr \citep{mukherjeeFirstMeasurementHubble2020,chenStandardSirenCosmological2022}, as well as for the possible association between GW190803 and J120437.98+500024.0 in \citet{heTracingLightIdentification2025}. Figure~\ref{fig:H0} shows the individual and combined constraints on the Hubble constant derived from these two candidate BBH--AGN associations. More generally, future samples of BBH mergers associated with flaring AGNs could significantly improve cosmological constraints \citep{bomStandardSirenCosmology2024}. %A recent analysis combining 13 candidate BBH--AGN associations reported an $H_0$ precision of approximately 4.4\%, highlighting the potential of multi-event analyses for standard-siren cosmology \citep{kumarHubbleConstantMeasurement2026a}.

Counterpart detections could also provide important constraints on BBH formation channels. A growing sample of mergers associated with AGNs would make it possible to estimate the contribution of the AGN disk channel to the overall BBH population \citep{bartosGravitationalwaveLocalizationAlone2017,zhuEvidenceFractionLIGO2025,zhuConstrainingFractionLIGO2026}. Combining host information with GW measurements of component masses, mass ratios, and spins could further test whether the properties of AGN-associated mergers are consistent with the population signatures predicted for this formation pathway \citep{yangHierarchicalBlackHole2019}.

EM counterparts could further probe the physical conditions in AGN disks. The luminosity, duration, and delay of a flare depend on the local disk environment, merger location, and recoil velocity, allowing observations of a counterpart to constrain these uncertain properties \citep{mckernanRampressureStrippingKicked2019,grahamCandidateElectromagneticCounterpart2020}. Joint GW and EM observations could therefore provide complementary information about the environments and physical processes associated with BBH mergers in AGN disks.

A secure EM counterpart would transform a BBH merger from an isolated GW source into a system with an identified astrophysical environment, enabling a much broader range of physical and cosmological studies.

\section{Summary and outlook}

The advent of multimessenger gravitational-wave astronomy has opened a new era in which compact-object mergers are no longer studied through gravitational waves alone, but are increasingly probed through coordinated searches for electromagnetic, neutrino, and other cosmic messengers. Within this broad context, this review has focused on one of the most challenging and observationally elusive questions in the field: whether BBH mergers occurring in the accretion disks of AGNs can produce detectable EM counterparts. In this review, we provide a systematic overview of the major advances in this rapidly developing field, with particular emphasis on three key questions. First, what are the formation channels and population signatures of BBH mergers in AGN disks, and how do they compare with other formation pathways? Second, what physical mechanisms can produce detectable EM counterparts, and what multi-wavelength signatures are predicted? Third, what is the current observational status of counterpart searches, and what constraints have been placed on the AGN contribution to the BBH population? By addressing these questions, we aim to clarify both the theoretical promise and the observational challenges of this emerging field.

Section 2 reviewed the formation channels and population signatures relevant to this question. BBH formation is commonly divided into isolated binary evolution, dynamical assembly and hierarchical mergers in dense stellar systems, and the AGN-disk channel. The AGN channel is distinctive because gas capture, migration, binary hardening, accretion, spin alignment, and repeated mergers can operate simultaneously. It is therefore expected to produce a high-mass tail, an unequal-mass subpopulation, high spin magnitudes, spin-alignment-dependent distributions of $\chi_{\rm eff}$ and $\chi_p$, and correlations among mass, mass ratio, and spin. More importantly, BBH systems formed through the AGN channel are expected to exhibit distinctive signatures in multiple source parameters simultaneously, such that their correlated features across different parameters can provide strong evidence for an AGN origin. Currently, hierarchical Bayesian inference and phenomenological mixture models have found evidence for a high-mass, high-spin subpopulation and a possible transition above $\sim 45~M_{\sun}$ consistent with hierarchical growth. Spatial-correlation studies have placed stringent upper limits on the contribution of luminous AGNs, while lower-luminosity and lower-Eddington-ratio AGNs may show preliminary positive correlations. In addition, significant spatial correlation between BBHs and AGNs exhibiting anomalous flares is also discovered in analysis. Both finds hint the AGN channel origins of some BBH events.

Section 3 summarized the theoretical models. The two main scenarios are in situ accretion onto the post-merger remnant and recoil-driven interaction of the remnant with the AGN disk. Both rely on hyper-Eddington accretion, which can power a wide-angle wind or a relativistic jet. The jet may break out of the disk, form a cocoon, and produce thermal and non-thermal emission. Typical predicted signals include soft X-ray transients, optical/ultraviolet flares, and radio afterglows, with luminosities, durations, and time delays depending strongly on uncertain disk and outflow parameters. In any case, future multi-wavelength observations, combined with theoretical modeling, will provide decisive evidence for the AGN channel of BBH formation.

Section 4 reviewed the observational searches. Early O1/O2 follow-up of GW150914 and other BBH events yielded no secure counterpart. The proposed GW190521-ZTF19abanrhr association marked a turning point, but its significance and physical origin remain debated. Subsequent AGN-targeted searches using ZTF, DECam, WFST, and other facilities have identified candidate flares associated with events such as GW231123 and GW190412. At present, the observational status is best described as theoretically compelling, statistically suggestive, and not yet conclusive. In our view, a key priority in the near term is to determine whether the AGN hosts of the currently known candidates will exhibit the secondary optical flares predicted by theoretical models. Compared with the initial optical flare, the secondary flare is expected to occur on a timescale of years. Once such a secondary flare is detected, prompt coordinated observations across the electromagnetic spectrum, from high-energy to radio bands (in particular for the radio band), should be initiated to provide critical evidence for further confirming its origin.

Taken together, the formation channels and population signatures, the theoretical models, and the observational searches demonstrate both the promise and the current limitations of this field. While no unambiguous EM counterpart to a BBH merger in an AGN disk has yet been secured, the field has matured rapidly, moving from speculative model building to systematic population inference and dedicated AGN-targeted searches. With improving gravitational-wave localization, expanding time-domain surveys, deeper multiwavelength follow-up, and more sophisticated models, the next decade offers a realistic prospect of identifying the first secure counterpart. Such a discovery would transform BBH mergers from isolated gravitational-wave sources into localized, cosmologically useful, and environmentally characterized multimessenger systems.

%%%====================================================
\section*{Acknowledgements}

The authors thank Zhengyan Liu, Rui Niu, Ning Jiang, Zhenyi Cai, Jian Li, Ji'an Jiang and Tao An for helpful discussions. This work is supported by the National Natural Science Foundation of China (grant Nos. 12325301 and 12405075), Strategic Priority Research Program of the Chinese Academy of Science (grant No.
XDB0550300), the National Key R\&D Program of China
(grant Nos. 2022YFC2204602 and
2024YFC2207500).

%%%====================================================
%%%====================================================
\end{CJK*}

\bibliography{ref}
\end{document}